\documentclass[11pt,a4paper]{article}
\usepackage{jheppub}
\usepackage[titletoc,toc,title]{appendix}

\title{Gravitational Waves from Phase Transitions in Scale Invariant Models}

\author[1]{Amine Ahriche$^{\href{https://orcid.org/0000-0003-0230-1774}{\includegraphics[width=2.5mm]{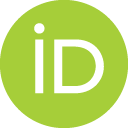}}}$}
\emailAdd{ahriche@sharjah.ac.ae}
\author[2]{Shinya Kanemura$^{\href{https://orcid.org/0000-0002-1303-7043}{\includegraphics[width=2.5mm]{ORCID.png}}}$}
\emailAdd{kanemu@het.phys.sci.osaka-u.ac.jp}
\author[2]{Masanori Tanaka$^{\href{https://orcid.org/0000-0002-1303-7043}{\includegraphics[width=2.5mm]{ORCID.png}}}$}
\emailAdd{m-tanaka@het.phys.sci.osaka-u.ac.jp}

\affiliation[1]{Department of Applied Physics and Astronomy, University of Sharjah, P.O. Box 27272 Sharjah, UAE.}
\affiliation[2]{Department of Physics, Osaka University, Toyonaka, Osaka 560-0043,
Japan.}

\abstract{We investigate the properties of the gravitational waves (GW) generated
during a strongly first order electroweak phase transition (EWPT) in
models with the classical scale invariance (CSI). Here, we distinguish
two parameter space regions that correspond to the cases of (1)
light dilaton and (2) purely radiative Higgs mass (PRHM). In the CSI models,
the dilaton mass, or the Higgs mass in the PRHM case, in
addition to some triple scalar couplings are fully triggered by the
radiative corrections (RCs). In order to probe the RCs effects on
the EWPT strength and on the GW spectrum, we extend the standard model
by a real singlet to assist the electroweak symmetry breaking and
an additional scalar field $Q$ with multiplicity $N_Q$ and mass $m_Q$.
After imposing all theoretical and experimental constraints, we show that a strongly first order EWPT with detectable GW spectra can be realized for the
two cases of light dilaton and PRHM. We also show the corresponding values of
the relative enhancement of the cross section for the di-Higgs production
process, which is related to the triple Higgs boson coupling.
We obtain the region in which the GW spectrum can be observed by
different future experiments such as LISA and DECIGO. We also show that the scenarios (1) and (2) can be discriminated by future GW observations and measurements of the di-Higgs productions at future colliders.}

\begin{document}

\begin{flushright}
{
OU-HET-1201
}
\end{flushright}

\maketitle

\section{Introduction}

The standard model (SM) in particle physics is a successful theory
to explain results at the Large Hadron Collider (LHC)~\cite{ATLAS:2012yve}.
On the other hand, the SM has several experimental and theoretical
problems. For the theoretical aspect, radiative corrections (RCs) to the Higgs boson mass cause the quadratic divergences in
the SM, which is the origin of so-called the hierarchy
problem~\cite{Bardeen:1995kv}. Related to this problem, extended
Higgs models with the classical scale invariance (CSI) have been often considered~\cite{Lee:2012jn,Englert:2013gz,Guo:2014bha,Endo:2015ifa,Hashino:2015nxa,Hashino:2016rvx,Ahriche:2016cio,Ahriche:2016ixu,Lane:2019dbc,Kanemura:2020yyr,Braathen:2020vwo,Ahriche:2021frb,Soualah:2021xbn, Eichten:2022vys}.
The CSI requires that all mass terms in the Lagrangian
are forbidden. Therefore, the electroweak symmetry breaking (EWSB)
does not occur at the tree-level. However, the CSI is violated by quantum effects of new particles. Then, the EWSB can be realized
radiatively. This mechanism is often called as the Coleman-Weinberg mechanism~\cite{Coleman:1973jx}. An extension of
the Coleman-Weinberg mechanism to the case with multi-scalar fields
has been also discussed by Gildener and Weinberg~\cite{Gildener:1976ih}.
The phenomenology of CSI extensions of the SM has been investigated
in the literature, including the testability at current and future
collider experiments~\cite{Lee:2012jn,Englert:2013gz,Guo:2014bha,Endo:2015ifa,Hashino:2015nxa,Hashino:2016rvx,Ahriche:2016cio,Ahriche:2016ixu,Lane:2019dbc,Kanemura:2020yyr,Braathen:2020vwo,Ahriche:2021frb,Soualah:2021xbn}.

Among the SM unsolved questions, the baryon asymmetry of the Universe
is an important problem for both particle physics and cosmology. The
observed baryon asymmetry is expressed by~\cite{Planck:2018vyg}
\begin{align}
\eta_{B}=\frac{n_{B}-n_{\bar{B}}}{n_{\gamma}}=6.143\pm0.190\times10^{-10},
\end{align}
where $n_{B}$, $n_{\bar{B}}$ and $n_{\gamma}$ refer to the number
density of baryons, antibaryons and the CMB photons, respectively.
In order to explain the ratio $\eta_{B}$, three conditions
need to be satisfied simultaneously at the early Universe, known as the Sakharov conditions~\cite{Sakharov:1967dj}.
These conditions can be summarized by the existence of such interactions
that violate the baryon number, violate C and CP symmetries;
and occur out of the thermal equilibrium. One of the most interesting
scenarios for baryogenesis is the electroweak baryogenesis (EWB),
where the third Sakharov condition is satisfied via
a strongly first order phase transition at the electroweak scale~\cite{Kuzmin:1985mm}.
However, it has been shown that this scenario cannot be realized in
the SM since the CP violation source in Yukawa interactions
is too small to produce $\eta_{B}$ and the electroweak phase transition
(EWPT) is not first order. Therefore, going beyond the SM is mandatory.
In order to realize the EWB scenario, the sphaleron processes should
decouple after the first order EWPT. This condition can be approximately
expressed by~\cite{Bochkarev:1987wf,Ahriche:2007jp,Ahriche:2014jna,Fuyuto:2014yia,Kanemura:2022ozv}
\begin{align}
\frac{\upsilon_{c}}{T_{c}}>1,\label{eq:vcTc}
\end{align}
where $T_{c}$ is the critical temperature at which the potential
minima (the false and true ones) are degenerate; and $\upsilon_{c}$
is the vacuum expectation value (VEV) of the $SU(2)_{L}$
doublet scalar field at $T=T_{c}$. Once the condition~\eqref{eq:vcTc} is satisfied,
the triple Higgs boson coupling $hhh$ significantly
deviates from the SM prediction~\cite{Kanemura:2004ch}. Therefore,
precise measurements of the $hhh$ coupling at future colliders are
important to test the scenario of the EWB. In order
to realize a large deviation in the triple Higgs boson coupling,
large quantum corrections from new particles play an important role.
Such large quantum corrections may be able to appear in the
other Higgs boson couplings like $h\gamma\gamma$~\cite{Shifman:1979eb, Kanemura:2002vm, Kanemura:2004mg, Arhrib:2012ia, Kanemura:2019kjg, Kanemura:2019slf, Degrassi:2023eii, Aiko:2023nqj}.

In addition to the $hhh$ coupling measurement, gravitational
waves (GWs) generated during a strongly first order EWPT can be
used to explore new physics models~\cite{Kamionkowski:1993fg,Kosowsky:1992rz,Kosowsky:1991ua,Schwaller:2015tja}.
A first order phase transition at the early Universe occurs via
the nucleation and expansion of bubbles of the broken
vacuum. When the broken vacuum bubbles collide with each other,
detectable GWs can be produced.
For a first order EWPT, the typical peak frequency of the GW spectrum
is around $10^{-3}-10^{-1}$\,Hz~\cite{Grojean:2006bp}. Such GW
spectrum can be observed by future space-based interferometers like
LISA~\cite{LISA:2017pwj} and DECIGO~\cite{Kawamura:2006up}. It
means that new physics models can be explored by using the GWs. Predictions
on the GW spectrum in extended Higgs models with the CSI have been discussed in the
literature~\cite{Kakizaki:2015wua, Huang:2016cjm, Hashino:2016rvx, Hashino:2016xoj, Hashino:2018zsi, Brdar:2018num, Brdar:2019qut, Salvio:2023qgb, Salvio:2023ynn}. The dynamics of the EWPT in the CSI models has been also discussed in the literature~\cite{Farzinnia:2014yqa, Mohamadnejad:2019vzg, Kubo:2015joa, Sannino:2015wka, Hashino:2016rvx, Kierkla:2022odc}.

Recently, it is often discussed that the first order EWPT can be tested
by primordial black hole observations~\cite{Hashino:2021qoq, Hashino:2022tcs}.
When the first order EWPT forms primordial black holes (PBHs), those
masses are around $10^{-5}$ times of the solar mass. It implies that
we can test the first order EWPT by observing such PBHs at current and future microlensing
experiments like Subaru HSC~\cite{Subaru}, OGLE~\cite{OGLE}, PRIME~\cite{PRIME} and Roman telescope~\cite{Roman}.

The CSI implies that a CP-even scalar should be massless
at tree-level; and becomes massive due to the quantum corrections that trigger the
EWSB. In Ref.~\cite{Ahriche:2021frb}, a generic model has been considered
to investigate the RC effects on the EWSB, dilaton
mass, scalar mixing, as well other observables. In this model, the
SM is extended by a real singlet and an extra singlet $Q$ with multiplicity
$N_{Q}$ and the couplings ($\alpha_{Q},\,\beta_{Q}$) to the Higgs
doublet and real singlet, respectively. Then, the RCs are quantified
by using $N_{Q}$, $\alpha_{Q}$ and $\beta_{Q}$ in addition to the
singlet VEV. In a case of large ($N_{Q}$, $\alpha_{Q}$, $\beta_{Q}$),
the dilaton mass can be 125~GeV due to the large RCs. This scenario
corresponds to the case of a purely radiative Higgs mass (PRHM)~\cite{Ahriche:2021frb}. Although large RCs (with large values for $N_{Q}$, $\alpha_{Q}$, and $\beta_{Q}$) are beneficial for vacuum stability, they may introduce tension with the triviality bound~\cite{Alexander-Nunneley:2010tyr,Kanemura:2012hr}. This aspect needs to be investigated within the viable parameter space of this model.

In this work, we will investigate the effect of quantum corrections on the EWPT dynamics and on the corresponding GW spectrum using this generic CSI model with
a scalar mixing~\cite{Ahriche:2021frb}. Here, the triple scalar
couplings are very sensitive to the quantum corrections, which makes the di-Higgs
production cross section at both LHC and ILC useful observables to
probe the viable parameter space. According to the full LHC Run 2
data with $139\,{\rm fb}^{-1}$, it is required that the cross section
for the non-resonant di-Higgs production process is
lower than 3.4 times of the SM prediction~\cite{CMS:2022dwd}. At
the High Luminosity LHC (HL-LHC), the cross section will be measured
by 0.7 times of the SM prediction~\cite{CMS:2022dwd,Cepeda:2019klc}.
At the International Linear Collider (ILC), it is expected that the
di-Higgs production cross section is limited less than 2 or 3 times
of the SM prediction~\cite{Bambade:2019fyw}.
The di-Higgs production processes in extended Higgs models have been investigated in
the literatures~\cite{Arhrib:2009hc, Dolan:2012ac, Kanemura:2016lkz, Dawson:2017jja, Carena:2018vpt, Arco:2021bvf, Abouabid:2021yvw, Iguro:2022fel}.
In this paper, we show
that the light dilaton and PRHM cases can be distinguished by using
complementarity between collider experiments and GW observations.

The structure of this paper is as follows. In Section~\ref{sec:CSI},
we introduce the CSI models; and discuss the EWSB and the scalar
mass spectrum. In Section~\ref{sec:constraints}, we identify the
theoretical and experimental constraints on these CSI models. In Section~\ref{sec:GW},
we define the parameters characterizing the first order EWPT and the
corresponding GW spectrum. Our numerical results for GW observations
and collider experiments are shown in Section~\ref{sec:results}.
Our conclusion is given in Section~\ref{sec:conclusion}.

\section{Models with Classical Scale Invariance \label{sec:CSI}}

The CSI in quantum field theories is defined by $x^{\mu}\to\kappa^{-1}x^{\mu},\,\psi_{i}\to\kappa^{a}\psi_{i}$ with $a=1,\,3/2$ for bosons and fermions, respectively. This invariance
implies that the scalar quadratic terms vanish. Therefore, the general
representation of the potential with the CSI is given by~\cite{Gildener:1976ih}
\begin{equation}
V(\Phi_{i})=\sum_{i,j,k,l}\lambda_{ijkl}\Phi_{i}\Phi_{j}\Phi_{k}\Phi_{l},\label{eq:V0}
\end{equation}
where $\Phi_{i}$ represents all the scalar representations. Obviously,
the EWSB could not take place because the scalar potential in Eq.~\eqref{eq:V0}
contains only quartic terms, which requires the contributions of the
quantum corrections that break both of the electroweak and CSI symmetries
at the same time. In order to achieve the EWSB in this class of models,
the SM is extended by a real singlet $S$; in addition to other bosonic
and fermionic degrees of freedom (dof) with $(n_{i},\,\alpha_{i},\,\beta_{i})$
as multiplicities and couplings to the Higgs doublet and scalar $S$,
respectively. Here, the Higgs doublet field ${\cal H}$ and the singlet
field $S$ are written as
\begin{equation}
{\cal H}=\left(\begin{array}{c}
\chi^{+}\\
\frac{H+i\,\chi^{0}}{\sqrt{2}}
\end{array}\right),\quad S=s,
\end{equation}
with $\chi^{+}$ and $\chi^{0}$ are the Goldstones; and $\langle H\rangle=\upsilon$
and $\langle s\rangle=\upsilon_{S}$ are the VEV of the doublet and
singlet, respectively. In this model, two CP-even eigenstates
$h_{1,2}$ (with $m_{2}>m_{1}$) are obtained by using the $2\times2$
mixing matrix with angle $\alpha$. We distinguish two possibilities
where the observed SM-like Higgs with the mass $m_{h}=125$ GeV corresponds
to (1) $h_{2}\equiv h$ and $h_{1}\equiv\eta$ is a light dilaton
(light dilaton scenario); and (2) $h_{1}\equiv h$ and $h_{2}\equiv\eta$
is a heavy scalar (PRHM scenario)~\cite{Ahriche:2021frb}.

The full one-loop effective potential in terms of CP-even fields $(H,s)$
can be written as
\begin{equation}
V^{1\ell}(H,s)=\frac{1}{24}(\lambda_{H}+\delta\lambda_{H})H^{4}+\frac{1}{24}(\lambda_{S}+\delta\lambda_{S})s^{4}+\frac{1}{4}(\omega+\delta\omega)H^{2}s^{2}+\sum_{i}n_{i}G\big(m_{i}^{2}(H,s)\big),\label{eq:V1l}
\end{equation}
with $\delta\lambda_{H},\delta\lambda_{S},\delta\omega$ are the counter
terms, $n_{i}$ and $m_{i}^{2}(H,s)$ are the multiplicities and the
field dependent masses for particles running loops, respectively.
Here, the function $G(r)$ is defined a la $\overline{DR}$ scheme,
i.e., $G(r)=\frac{r^{2}}{64\pi^{2}}\big(\log(r/\Lambda^{2})-3/2\big)$,
with the renormalization scale is taken to be $\Lambda=m_{h}=125.18\,\mathrm{GeV}$.

In this setup, we can eliminate the couplings $\lambda_{H},\lambda_{S}$
and $\omega$ in favor of the tadpole conditions and the Higgs mass
at tree-level, and the corresponding counter-terms $\delta\lambda_{H},\,\delta\lambda_{S},\,\delta\omega$
can be also eliminated using the one-loop tadpole
and Higgs mass conditions as shown in Refs.~\cite{Ahriche:2021frb,Soualah:2021xbn}.
In case where all field dependent masses can be written in the form\footnote{Indeed, all field dependent masses can be written in this form except
the eigenmasses of $h_{1,2}$. The contributions of $h_{1,2}$ to
the effective potential can be safely neglected.} $m_{i}^{2}(H,s)=\frac{1}{2}(\alpha_{i}H^{2}+\beta_{i}s^{2})$, the
counter-terms can be simplified as
\begin{align}
\delta\omega & =\frac{m_{h}^{2}}{\upsilon^{2}+\upsilon_{S}^{2}}\frac{(ab-c^{2})(\upsilon^{2}+\upsilon_{S}^{2})-a\upsilon_{S}^{2}-b\upsilon^{2}+2\,c\upsilon\upsilon_{S}}{a\upsilon^{2}+b\upsilon_{S}^{2}+2\,c\upsilon\upsilon_{S}-\upsilon^{2}-\upsilon_{S}^{2}},\label{eq:dw}
\end{align}
with
\begin{align}
\begin{aligned}a & =\frac{1}{32\pi^{2}m_{h}^{2}}\sum_{i}n_{i}\alpha_{i}\Big(2m_{i}^{2}-\beta_{i}\upsilon_{S}^{2}\log\frac{m_{i}^{2}}{\Lambda^{2}}\Big),\nonumber\\
b & =\frac{1}{32\pi^{2}m_{h}^{2}}\sum_{i}n_{i}\beta_{i}\Big(2m_{i}^{2}-\alpha_{i}\upsilon^{2}\log\frac{m_{i}^{2}}{\Lambda^{2}}\Big),\nonumber\\
c & =\frac{\upsilon\upsilon_{S}}{32\pi^{2}m_{h}^{2}}\sum_{i}n_{i}\alpha_{i}\beta_{i}\log\frac{m_{i}^{2}}{\Lambda^{2}}.
\end{aligned}
\end{align}

The light dilaton and PRHM cases are identified via the conditions~\cite{Ahriche:2021frb,Soualah:2021xbn}
\begin{align}
\delta\omega(\upsilon^{2}+\upsilon_{S}^{2})/m_{h}^{2}<a+b<1+\delta\omega(\upsilon^{2}+\upsilon_{S}^{2})/m_{h}^{2} & \quad(\mbox{for light dilaton case}),\label{dilaton}\\
a+b>1+\delta\omega(\upsilon^{2}+\upsilon_{S}^{2})/m_{h}^{2} & \quad(\mbox{for PRHM case}),\label{higgs}
\end{align}
respectively. However, the condition $a+b=1+\delta\omega(\upsilon^{2}+\upsilon_{S}^{2})/m_{h}^{2}$
corresponds to the special case of degenerate eigenmasses $m_{1}=m_{2}=m_{h}$.
This case is of great interest that deserves an independent study.

In our analysis, we consider a generic model where
the SM is extended by a scalar singlet $S$ to assist the EWSB; and
another boson $Q$ with multiplicity $N_{Q}$ and the squared mass
$m_{Q}^{2}=\frac{1}{2}(\alpha_{Q}\upsilon^{2}+\beta_{Q}\upsilon_{S}^{2})$.
Clearly, the quantum correction from the boson $Q$ is proportional
to the field multiplicity $N_{Q}$, the couplings ($\alpha_{Q},\,\beta_{Q}$)
to $\mathcal{H}$ and $S$ and/or the singlet VEV $\upsilon_{S}$.

\section{Constraints and Predictions~\label{sec:constraints}}

The different constraints on this model have been discussed in details
in Ref.~\cite{Ahriche:2021frb}. Here, we mention the constraints
coming from the total Higgs strength modifier $\mu_{{\rm tot}}=c_{\alpha}^{2}\times(1-\mathcal{B}_{BSM})\geq0.89$
at 95~\% CL~\cite{ATLAS:2016neq}, that implies the scalar mixing
to be $s_{\alpha}^{2}\leq0.11$ in the absence of invisible and undetermined
Higgs decay ($\mathcal{B}_{BSM}=0$). Here, the RCs play a crucial
role to satisfy this bound. The mixing sine can be decomposed into
a tree-level part and a one-loop part as $s_{\alpha}=s_{\alpha}^{(0)}+s_{\alpha}^{(1)}$,
where $s_{\alpha}^{(0)}=\upsilon/\sqrt{\upsilon^{2}+\upsilon_{S}^{2}}$
in the light dilaton case. Therefore, small RCs will keep the constraint
$s_{\alpha}^{2}\leq0.11$ fulfilled for $\upsilon_{S}\gg\upsilon$.
However, in the PRHM case, we have $s_{\alpha}^{(0)}=\upsilon_{S}/\sqrt{\upsilon^{2}+\upsilon_{S}^{2}}$
which makes the constraint $s_{\alpha}^{2}\leq0.11$ explicitly violated
for $\upsilon_{S}\gg\upsilon$. Interestingly, large RCs in the PRHM
case that are responsible to push the light CP-even eigenmass from zero tree-level value to the measured 125\,GeV; are
also responsible to generate large negative contributions to the mixing
$s_{\alpha}^{(1)}$ that makes the condition $s_{\alpha}^{2}\leq0.11$
fulfilled. This point has been discussed in details in Ref.~\cite{Ahriche:2021frb}
for this model; and in Ref.~\cite{Soualah:2021xbn} for the SI scotogenic
model.

The counter-terms $\delta\lambda_{H},~\delta\lambda_{S},~\delta\omega$
cannot take any numerical values since they are constrained by the
one-loop perturbativity conditions $\lambda_{H}^{1-\ell},~\lambda_{S}^{1-\ell},~|\omega^{1-\ell}|<4\pi$.
Here, the one-loop quartic couplings are defined as
the $4^{th}$ derivatives of the full one-loop scalar potential in
Eq.~\eqref{eq:V1l} at the broken vacuum. In what follows, we will
consider the one-loop value for the Higgs-dilaton mixing angle, since
it has been shown that the role of RCs is crucial in driving
the mixing value towards the experimentally allowed region, both in
the context of a light dilaton and PRHM cases~\cite{Ahriche:2021frb}.
In the CSI models, the leading term in the one-loop
scalar potential is $\varphi^{4}\log\varphi$ rather than $\varphi^{4}$,
where $\varphi$ could be any direction in the $H$-$s$ plane.
Thus, the vacuum stability conditions differ from those used in the literature,
which are given by $\sum_{i}n_{i}\alpha_{i}^{2}>0\,\wedge\,\sum_{i}n_{i}\beta_{i}^{2}>0$~\cite{Ahriche:2021frb,Soualah:2021xbn}.

In addition to the perturbativity and vacuum stability conditions, we discuss here the triviality bound as a theoretical constraint on our model.
The triviality bound plays a significant role in discussing upper bounds on couplings~\cite{Lindner:1985uk}.
Especially, the triviality bound can give strong constraints on models with the CSI or a strongly first order EWPT~\cite{Alexander-Nunneley:2010tyr, Kanemura:2012hr}. Generally, the Landau pole scale is defined as the energy scale at which the perturbative description of the model is broken down, which suggests the need for a more fundamental theory or modifications at higher energy scales.
In this work, we define the Landau pole scale $(\Lambda_{\rm Lan})$ as the approximate scale where $\Lambda_{\rm Lan}=\min(\{\Lambda_i\})$ with $\lambda_i(\mu=\Lambda_i)=4\pi$, and $\lambda_{i}(\mu)$'s are the couplings at the scale $\mu$ in our model. In what follow, we consider the condition $\Lambda_{\rm Lan} < 10\,{\rm TeV}$ as a triviality bound.

In Appendix~\ref{sec:Landau}, we present the $\beta$-functions for the renormalization group equations
(RGEs) in our model using the ordinary scheme. However, these functions do not decouple the effects of
heavy new particles in the RGEs flow, even when our focus is on physics at a low energy scale. These
non-decoupling effects are commonly known as threshold effects. Recently, it has been confirmed that
such threshold effects can be naturally taken into account by employing mass-dependent
$\beta$-functions~\cite{Kanemura:2023wap}. In the following, we will utilize the mass-dependent
$\beta$-function when discussing the RGEs analysis.

As well known, the triple Higgs boson coupling, $\lambda_{hhh}=\lambda_{hhh}^{{\rm SM}}(1+\Delta_{hhh})$,
is an important quantity to probe the strongly first order EWPT~\cite{Grojean:2004xa,Kanemura:2004ch,Noble:2007kk}.
In order to obtain information about the relative triple
Higgs coupling enhancement ($\Delta_{hhh}$), it is necessary to measure
the di-Higgs production processes at the LHC or the ILC. However,
in models where the SM Higgs doublet mixes with a singlet, as in our model,
the di-Higgs production process involves an extra Feynman triangle diagram
mediated by the extra scalar $\eta$.
In this case, the di-Higgs production cross section has three independent
contributions that come from: (1) the Feynman diagrams involving only
the triple scalar couplings $\lambda_{hhh}$ and $\lambda_{hh\eta}$
($\sigma_{\lambda}$), (2) the diagrams with only pure gauge couplings
($\sigma_{G}$); and (3) the interference contribution ($\sigma_{G\lambda}$)~\cite{Ahriche:2014cpa,Ahriche:2021frb}.
In our model, the di-Higgs production cross section
scaled by its SM value can be expressed as
\begin{align}
R(hh+X)=\frac{\sigma(hh+X)}{\sigma_{{\rm SM}}(hh+X)}=\frac{\xi_{1}\sigma_{G}+\xi_{2}\sigma_{\lambda}+\xi_{3}\sigma_{G\lambda}}{\sigma_{G}+\sigma_{\lambda}+\sigma_{G\lambda}}.\label{R}
\end{align}
The coefficients $\xi_{i}~(i=1,2,3)$ are defined at the CM energy
$\sqrt{s}$ as~\cite{Ahriche:2014cpa,Ahriche:2021frb}
\begin{align}
\xi_{1}=c_{\alpha}^{4},\quad\xi_{2}=|\mathcal{P}|^{2},\quad\xi_{3}=c_{\alpha}^{2}\Re(\mathcal{P}),
\end{align}
with
\begin{align}
\mathcal{P}=c_{\alpha}\frac{\lambda_{hhh}}{\lambda_{hhh}^{{\rm SM}}}+s_{\alpha}\frac{\lambda_{hh\eta}}{\lambda_{hhh}^{{\rm SM}}}\frac{s-m_{h}^{2}+im_{h}\Gamma_{h}}{s-m_{\eta}^{2}+im_{\eta}\Gamma_{\eta}},
\end{align}
where $\Gamma_{h}$ is the measured Higgs total decay width, $\Gamma_{\eta}$
is the estimated heavy scalar total decay width; and $\lambda_{hhh}^{{\rm SM}}$
is the triple Higgs boson coupling in the SM. We take the value of
$\lambda_{hhh}^{{\rm SM}}$ as in Refs.~\cite{Kanemura:2002vm,Kanemura:2004mg}.

Here, we consider the di-Higgs production processes $pp\to hh$ and
$e^{+}e^{-}\to Zhh$ at the LHC with 14\,TeV and ILC with 500\,GeV,
respectively. In the following, we use the notations $R_{{\rm LHC}}$
and $R_{{\rm ILC}}$ for the processes $R(pp\to hh{\rm @LHC14\,TeV})$
and $R(e^{+}e^{-}\to Zhh{\rm @ILC500\,GeV})$, respectively. The values
$\sigma_{\lambda},~\sigma_{G}$ and $\sigma_{G\lambda}$ for the process
$pp\to hh{\rm @LHC14\,TeV}$ and $e^{+}e^{-}\to Zhh{\rm @ILC500\,GeV}$
are given in Table.~\ref{tab:sigma}.
\begin{table}[t]
\centering %
\begin{tabular}{|c|c|c|c|c|}
\hline
 & $\sigma_{\lambda}$ [fb] & $\sigma_{G}$ [fb] & $\sigma_{G\lambda}$ [fb] & Reference \tabularnewline
\hline
~$pp\to hh\,@{\rm LHC}\,14$TeV~ & ~70.1~ & ~9.66~ & ~$-49.9$~ & \cite{Spira:1995mt} \tabularnewline
\hline
~$e^{+}e^{-}\to Zhh\,@{\rm ILC}\,500$GeV~ & ~0.0837~ & ~0.01565~ & ~0.05685~ & \cite{Ahriche:2021frb} \tabularnewline
\hline
\end{tabular}\caption{ The cross section contributions in Eq.~\eqref{R} for each di-Higgs
production process at the LHC and the ILC.}
~\label{tab:sigma}
\end{table}

One has to notice that, in the PRHM scenario, the heavy
scalar $\eta$ decays into all SM Higgs final states in addition to
a di-Higgs channel. This means that the negative searches for a
heavy resonance at both ATLAS and CMS can be used to constraint our
model, like: (1) the heavy CP-even resonance in the channels of pair
of leptons, jets or gauge bosons $pp\to H\to\ell\ell,\,jj,\,VV$~\cite{ATLAS:2020zms,ATLAS:2020tlo,CMS:2021klu};
and (2) the resonance in the di-Higgs production $pp\to H\to hh$~\cite{ATLAS:2021fet,ATLAS:2021ulo,ATLAS:2021jki}.
It has been shown in Ref.~\cite{Ahriche:2021frb} that
almost the full parameter space region is allowed even if we consider the
heavy CP-even resonance. Thus, these constraints will not be considered
in our work.

\section{Gravitational Waves from a First Order Phase Transition~\label{sec:GW}}

In order to analyze the GW spectrum generated during a strongly
first order EWPT, we need to define the parameters $\alpha$ and $\beta$,
which characterize the GWs from the dynamics of vacuum bubbles~\cite{Grojean:2006bp}.
These parameters represent the latent heat and the inverse of the
EWPT time duration, respectively. However, an accurate description
of the EWPT dynamics is mandatory, which requires the exact knowledge
of a key physical quantity: the full one-loop effective potential
at finite temperature\footnote{Recently, there has been a discussion on methods to decrease the
uncertainty associated with the thermal effective potential, considering factors such as renormalization
scale and gauge dependencies~\cite{Croon:2020cgk, Lofgren:2021ogg, Schicho:2022wty}. }, that is given
by~\cite{Dolan:1973qd}
\begin{align}
V_{{\rm eff}}^{T}(H,s,T) & = V^{1\ell}(H,s)+T^{4}\sum_{i={\rm all}}n_{i}J_{B,F}\big(m_{i}^{2}/T^{2}\big)+V_{{\rm daisy}}(H,s,T),\label{Veff}\\
J_{B,F}\left(\alpha\right) & = \frac{1}{2\pi^{2}}\int_{0}^{\infty}drr^{2}\log\left[1\mp\exp(-\sqrt{r^{2}+\alpha})\right],\label{JBF}
\end{align}
where the last contribution in Eq.~\eqref{Veff}, that is called
the daisy (or ring) part, represents the leading term of higher order
thermal corrections~\cite{Carrington:1991hz}. This contribution
can be taken into account by replacing the scalar and longitudinal
gauge field-dependent masses in the first and second terms of Eq.~\eqref{Veff} by their thermally corrected values,
i.e., $m_{i}^{2}\rightarrow\widetilde{m}_{i}^{2}=m_{i}^{2}+\Pi_{i}$~\cite{Gross:1980br}.
The thermal self-energies are given by~\cite{Ahriche:2018rao}
\begin{align}
\begin{aligned} & \Pi_{H}=\Pi_{\chi}=T^{2}\left(\frac{1}{2}\lambda_{H}+\frac{3}{16}g^{2}+\frac{1}{16}g^{\prime2}+\frac{1}{4}y_{t}^{2}+\frac{\omega}{24}+\frac{\alpha_{Q}N_{Q}}{24}\right),\\
 & \Pi_{S}=\Big(\frac{\omega}{6}+\frac{\lambda_{S}}{24}+\frac{\beta_{Q}N_{Q}}{24}\Big)\,T^{2},\\
 & \Pi_{Q}=T^{2}\left(\frac{\alpha_{Q}}{4}+\frac{\beta_{Q}}{12}+\frac{\lambda_{Q}N_{Q}}{24}\right),\\
 & \Pi_{W_{L}}=\frac{11}{6}g^{2}T^{2},\quad\Pi_{B_{L}}=\frac{11}{6}g^{\prime2}T^{2},\quad\Pi_{W_{L},B_{L}}\simeq0,
\end{aligned}
\end{align}
with $g$ and $g^{\prime}$ are the $SU(2)_{L}$ and $U(1)_{Y}$ gauge
couplings, respectively. Here, $\lambda_{Q}$ is the self-coupling
constant of boson $Q$. Since the thermal corrections related to $\lambda_{Q}$
only come from the thermal mass of the new boson $Q$, this correction
can be negligible. Thus, for reasons of simplicity, we take $\lambda_{Q}=0$
in our analysis.

The EWSB takes place during the transition from the symmetric vacuum
($\langle H\rangle=0$) to the broken one ($\langle H\rangle\ne0$),
at the nucleation temperature $T_{n}\le T_{c}$. A strongly first order EWPT
occurs through the tunneling between the symmetric
and broken vacuum, which corresponds to the vacuum bubble nucleation
at random points in the space. These bubbles expand and collide with
each other until filling the Universe by the broken vacuum ($\langle H\rangle\ne0$).
When a bubble wall passes through an unbroken symmetry region (where
$\langle H\rangle=0$) at which a net baryon asymmetry is generated
by the $B+L$ and CP violating processes, the thermal equilibrium will washout
this asymmetry unless the $B$ number violating process is suppressed inside
the bubble (broken vacuum $\langle H\rangle\neq0$).
This condition is often called as the sphaleron decoupling condition and approximately expressed by Eq.~\eqref{eq:vcTc}.
It has been shown that the singlet plays an important
role in the EWPT dynamics even though its VEV $\langle S\rangle_{c}$
is absent in the condition in Eq.~\eqref{eq:vcTc}~\cite{Ahriche:2007jp,Ahriche:2014jna}.
The precise evaluation of the sphaleron decoupling condition in extended Higgs models has been also discussed in Refs.~\cite{Fuyuto:2014yia,Kanemura:2022ozv}.

In what follows, we analyze the GW spectrum from a first order EWPT
in this model by estimating the above mentioned parameters $\alpha$
and $\beta$~\cite{Grojean:2006bp}. The parameter $\alpha$ is the
latent heat normalized by the radiation energy density, which is given
by
\begin{equation}
\alpha=\frac{\epsilon(T_{n})}{\rho_{{\rm rad}}(T_{n})},\label{eq:alpha}
\end{equation}
with $\rho_{{\rm rad}}(T)=(\pi^{2}/30)g_{\ast}T^{4}$ is the radiation
energy density, $g_{\ast}$ is relativistic degrees of freedom in
the thermal plasma; and
\begin{equation}
\epsilon(T)=-V_{{\rm eff}}^{T}(H(T),s(T),T)+T\frac{\partial V_{{\rm eff}}^{T}(H(T),s(T),T)}{\partial T},
\end{equation}
is the released energy density, where the configuration $(H(T),s(T))$
is the broken vacuum at temperature $T$.

The parameter $\beta$ that describes approximately the inverse of
time duration of the EWPT is defined as
\begin{equation}
\beta=-\left.\frac{dS_{E}}{dt}\right|_{t=t_{n}}\simeq\left.\frac{1}{\Gamma}\frac{d\Gamma}{dt}\right|_{t=t_{n}},\label{eq:beta}
\end{equation}
where $S_{E}$ and $\Gamma$ are the 4d Euclidean action of a critical
bubble and vacuum bubble nucleation rate per unit volume per unit
time at the time of the EWPT $t_{n}$, respectively. In the following,
we use the normalized parameter $\beta$ by Hubble
parameter $H_{T}$
\begin{equation}
\widetilde{\beta}=\frac{\beta}{H_{T}}=T_{n}\frac{d}{dT}\left(\frac{S_{3}(T)}{T}\right)\Bigg|_{T=T_{n}},
\end{equation}
where $S_{3}(T)$ is the 3d action for the bounce solution. The transition
temperature $T_{n}$ is defined by
\begin{equation}
\left.\frac{\Gamma}{H_{T}^{4}}\right|_{T=T_{n}}=1,\label{eq:TransComp}
\end{equation}
with the bubble nucleation rate~\cite{Linde:1981zj}
\begin{equation}
\Gamma(T)\simeq T^{4}\left(\frac{S_{3}(T)}{2\pi T}\right)^{3/2}\exp\left[-\frac{S_{3}(T)}{T}\right].
\end{equation}

If $\Gamma/H_{T}^{4}$ cannot be larger than the unity ($\Gamma/H_{T}^{4}\ll1$),
the first order phase transition does not complete by today.
Thus, this condition is used as another theoretical constraint
on our model. Here, we use the public code \texttt{CosmoTransitions}
to obtain the bounce solutions~\cite{Wainwright:2011kj}.

The GWs from a first order phase transition can be produced via three
mechanisms: (1) collisions of bubble walls and shocks in the plasma
$\Omega_{\phi}h^{2}$~\cite{Kosowsky:1992rz}, (2) the compressional
waves (sound waves) $\Omega_{{\rm sw}}h^{2}$~\cite{Huber:2008hg};
and (3) magnetohydrodynamic (MHD) turbulence in the plasma $\Omega_{{\rm tur}}h^{2}$~\cite{Caprini:2006jb}.
Therefore, the stochastic GW background can be approximately expressed
by
\begin{equation}
\Omega_{{\rm GW}}h^{2}\simeq\Omega_{\phi}h^{2}+\Omega_{{\rm sw}}h^{2}+\Omega_{{\rm tur}}h^{2}.
\end{equation}

The importance of each contribution depends on the EWPT dynamics,
especially on the bubble wall velocity $v_{b}$.
In this work, we take $v_{b}=0.95$ as a free parameter;
and focus on the contribution to GWs from the sound waves in the plasma,
which is the dominant contributions among the other GW sources.
According to the numerical simulations, a fitting function for the
GW spectrum from the sound waves is given by~\cite{Caprini:2015zlo}
\begin{equation}
\Omega_{{\rm sw}}(f)h^{2}=\widetilde{\Omega}_{{\rm sw}}h^{2}\times(f/\tilde{f}_{{\rm sw}})^{3}\left(\frac{7}{4+3(f/\tilde{f}_{{\rm sw}})^{2}}\right)^{7/2},
\end{equation}
where the peak of the amplitude is given by
\begin{equation}
\widetilde{\Omega}_{{\rm sw}}h^{2}\simeq2.65\times10^{-6}v_{b}\widetilde{\beta}^{-1}\left(\frac{\kappa\alpha}{1+\alpha}\right)^{2}\left(\frac{100}{g_{\ast}}\right)^{1/3};
\end{equation}
and the peak frequency is expressed as
\begin{equation}
\widetilde{f}_{{\rm sw}}\simeq1.9\times10^{-5}\mbox{Hz}\frac{1}{v_{b}}\widetilde{\beta}\left(\frac{T_{n}}{100{\rm GeV}}\right)\left(\frac{g_{\ast}}{100}\right)^{1/6},
\end{equation}
with $\kappa$ is the efficiency factor which characterize how much
of the vacuum energy is converted into the fluid motion~\cite{Espinosa:2010hh}.

\section{Numerical Results~\label{sec:results}}

\begin{figure}[t]
\includegraphics[width=0.99\textwidth]{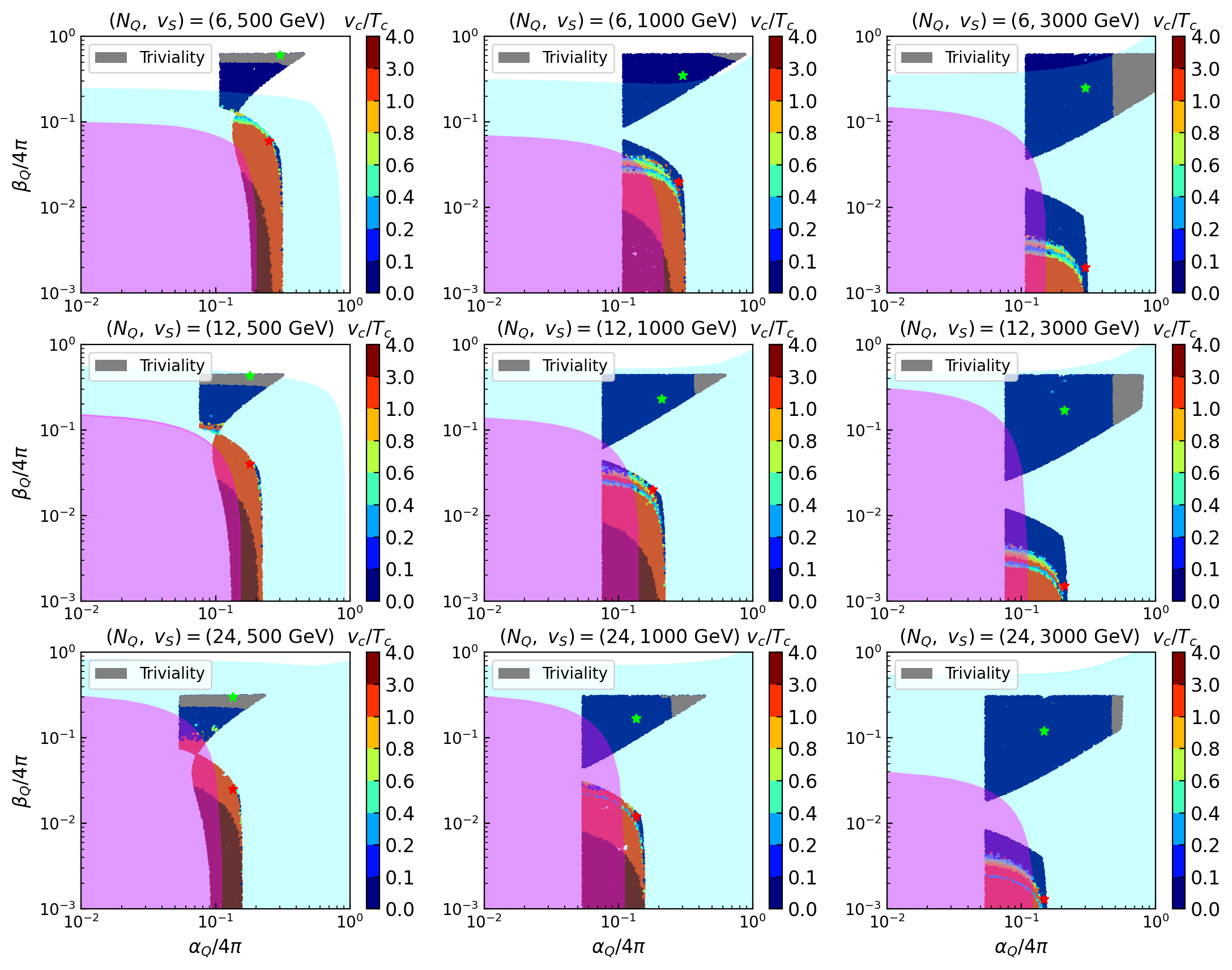}
\caption{The dependence of the EWPT strength parameter $\upsilon_{c}/T_{c}$
on the couplings ($\alpha_{Q},~\beta_{Q}$) with different values
of the multiplicity $N_{Q}$ and the singlet VEV $\upsilon_{S}$.
The green and red stars are the benchmark points (BPs)
that correspond to the PRHM and light dilaton scenarios, respectively,
which will be used in Fig.~\ref{fig:GWs}.
The area colored in cyan represents the parameter space region that could be tested by DECIGO, while the magenta region is eliminated by the completion condition
for the first order EWPT~\eqref{eq:TransComp}.
The gray region is constrained by the triviality bound with $\Lambda_{\rm Lan} < 10\,{\rm TeV}$.}
\label{fig:vcTc}
\end{figure}

In our analysis, we consider the following values for the multiplicity
$N_{Q}=6,\,12,\,24$; and the singlet VEV $\upsilon_{S}=500\,\mathrm{GeV},~1\,\mathrm{TeV},~3\,\mathrm{TeV}$,
while allowing the couplings to lie within the perturbative
regime $|\alpha_{Q}|,|\beta_{Q}|\leq4\pi$. In addition,
we consider the constraints: (1) vacuum stability, (2) the mixing
angle $s_{\alpha}^{2}\leq0.11$, (3) the perturbativity one-loop constraints
$\lambda_{H,S}^{1-\ell},\,|\omega^{1-\ell}|<4\,\pi$; and (4) the
completion condition of the phase transition. As well known,
the GWs from a first order EWPT is useful to probe the structure
of the Higgs potential~\cite{Kakizaki:2015wua,Hashino:2016rvx}.
Since the RCs have the key role in this model, one expects that future
space-based interferometers would be able to detect the GWs generated
during the first order EWPT. In order to understand the impact
on the EWPT strength from the quantum corrections, we show the ratio
$\upsilon_{c}/T_{c}$ in the palette in Fig.~\ref{fig:vcTc} for different values of
$(\alpha_{Q},~\beta_{Q})$, $N_{Q}$ and $\upsilon_{S}$.

In all panels of Fig.~\ref{fig:vcTc}, the upper parameter space region
corresponds to the PRHM scenario; and the lower region is the light
dilaton case. The regions colored in cyan represent the parameter
region where the GW spectra from the first order EWPT may be detected
at DECIGO~\cite{Kawamura:2006up}. Here, one notices
that not only the dilaton case, but also the PRHM scenario predicts
testable GW spectra. The magenta region in Fig.~\ref{fig:vcTc}
is constrained by the completion condition of the first order EWPT.
Interestingly, in Fig.~\ref{fig:vcTc}, there is a parameter space region
where the GW spectra can be tested at future space-based interferometers
even if the EWPT is not strongly first order, i.e., the condition
in Eq.~\eqref{eq:vcTc} is not satisfied.
One remarks that satisfying the condition in Eq.~\eqref{eq:vcTc} requires
small $\alpha_{Q}$ and $\beta_{Q}$. Furthermore, the triviality bound ($\Lambda_{\rm Lan}<10~\rm{TeV}$)
excludes a portion of the parameter space corresponding to large values of $\alpha_{Q}$ and
$\beta_{Q}$ in the PRHM case. This observation has also been noted in extended scalar sector models,
where more parameter space is excluded when the triviality bound is considered at a higher scale
~\cite{Khojali:2022squ,Baouche:2021wwa}. On the other hand, as we will
explain later, the di-Higgs production cross section can be large
for relatively large $\beta_{Q}$ value. It indicates a complementarity
between the GW observations and collider experiments to test our
model. The details for the di-Higgs production will be discussed
later.

According to Fig.~\ref{fig:vcTc}, as $N_{Q}$ is
getting large, the parameter space region satisfying the condition in~\eqref{eq:vcTc}
becomes large in the PRHM scenario. Whereas, in the light dilaton
case, such parameter space region is narrowed down, which means that the
PRHM scenario is preferred in order to realize the EWBG scenario within
CSI models. From the results in Fig.~\ref{fig:vcTc}, one learns that the
scenario with degenerate masses $m_{1}=m_{2}=m_{h}$ is possible;
and it corresponds to the interference region between the upper and
lower islands. This option cannot be realized in cases with the large
$\upsilon_{S}$.

\begin{figure}[t]
\includegraphics[width=0.99\textwidth]{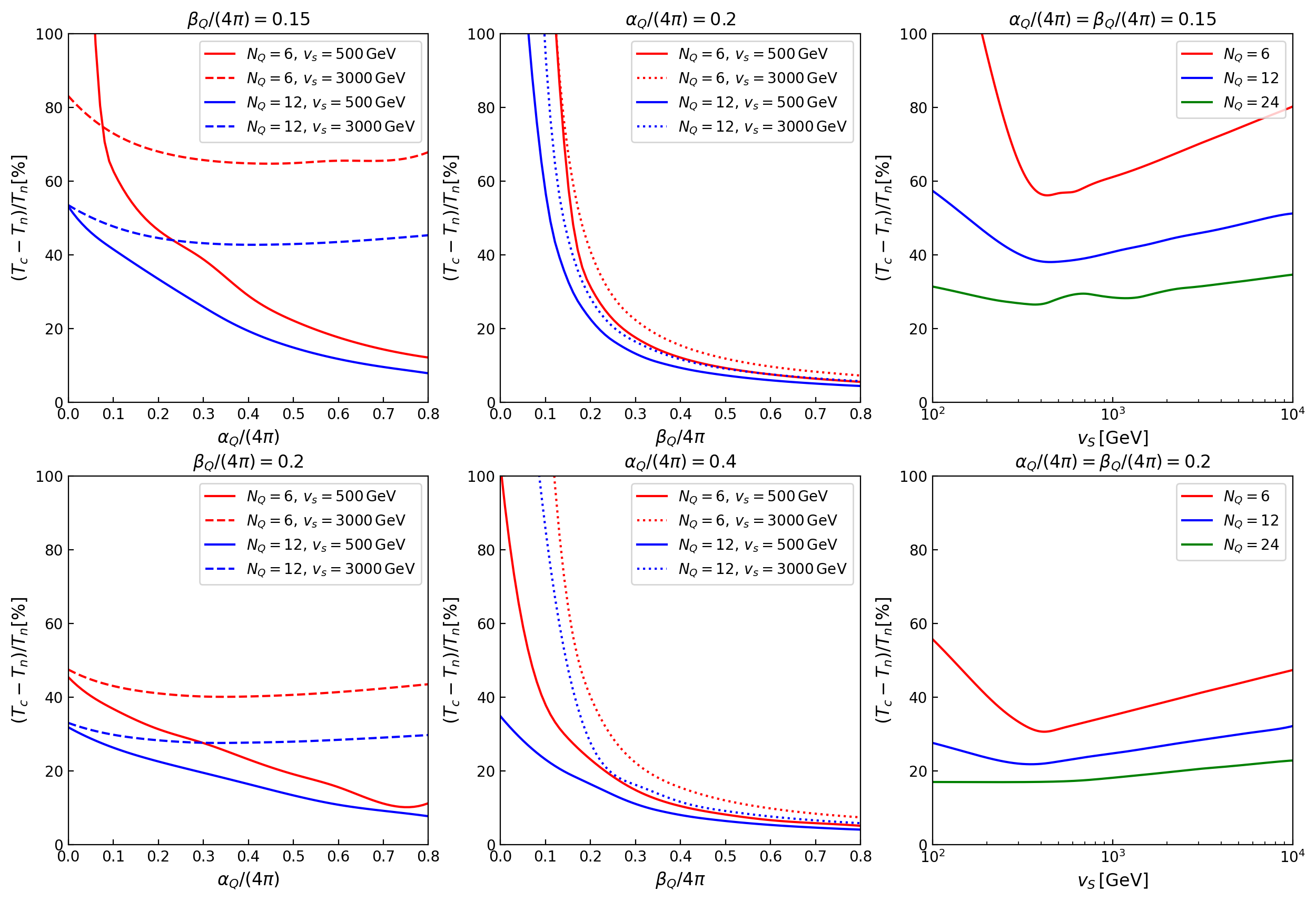}
 \caption{The relative difference between the critical and the nucleation temperature
values in function of the couplings $\alpha_{Q}$ (left) $\beta_{Q}$
(middle) and the singlet VEV $\upsilon_{S}$ (right). The phase transition
in our model is clearly supercooling. }
\label{fig:Tc_Tn}
\end{figure}

In order to understand more about the EWPT dynamics, we show the relative
difference between the critical and the nucleation temperature
values $(T_{c}-T_{n})/T_{n}$ in Fig.~\ref{fig:Tc_Tn} for different
values of the model free parameters.

Since this difference is significantly large, the
first order EWPT in our analysis can be supercooling. In Fig.~\ref{fig:Tc_Tn},
it is shown that the relative difference $(T_{c}-T_{n})/T_{n}$ is
large for larger singlet VEV $\upsilon_{S}$, and/or smaller couplings
($\alpha_{Q},\,\beta_{Q}$). As a consequence, the testable GWs can be produced
even if the EWPT strength parameter $\upsilon_{c}/T_{c}$ is smaller than unity.
This is the reason why the wide parameter space region in
Fig.~\ref{fig:vcTc} can be probed by GW experiments.

By considering the two BPs that correspond to the green (PRHM case) and red (light dilaton case) stars in Fig.~\ref{fig:vcTc},
we show the GW amplitude $\Omega_{{\rm GW}}h^{2}$ as functions of
the frequency values.
\begin{figure}[t]
\includegraphics[width=0.99\textwidth]{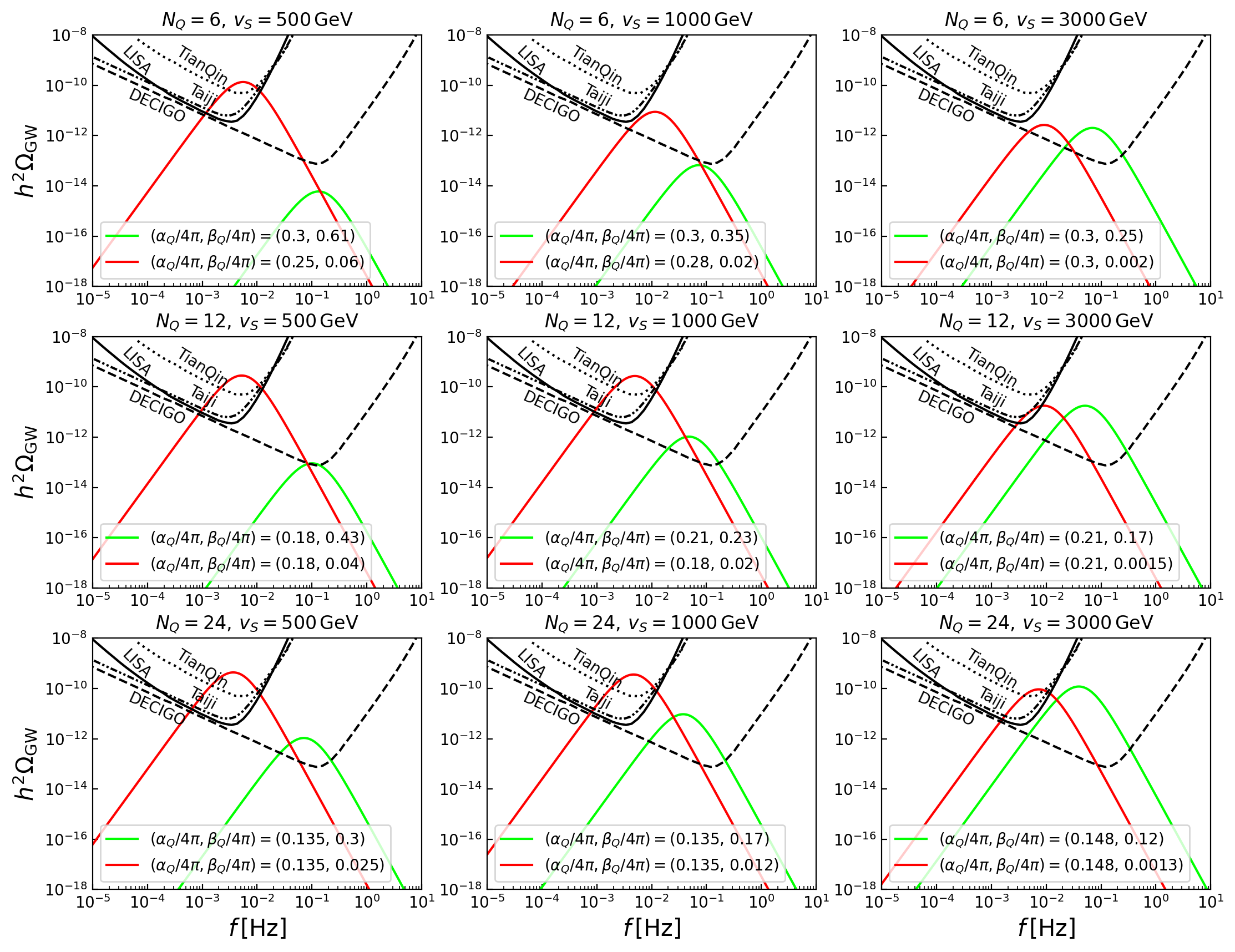} \caption{The predicted GW spectra for the two
BPs with different values of the multiplicity $N_{Q}$ and the singlet VEV $\upsilon_{S}$.
The red and green lines correspond to the light dilaton
scenario and the PRHM scenario, respectively, as shown as the green
and red starts in Fig.~\ref{fig:vcTc}. We also show the sensitivity
curves of LISA~\cite{LISA:2017pwj}, DECIGO~\cite{Kawamura:2006up},
TianQin~\cite{TianQin:2015yph} and Taiji~\cite{Ruan:2018tsw}.}
\label{fig:GWs}
\end{figure}
We conclude from Fig.~\ref{fig:GWs} that our model
may be tested at future space-based interferometers such as LISA~\cite{LISA:2017pwj},
DECIGO~\cite{Kawamura:2006up}, TianQin~\cite{TianQin:2015yph}
and Taiji~\cite{Ruan:2018tsw}. Even if the predicted GW spectrum
cannot reach the sensitivity curves of the future GW observations,
we may be able to extract the information of the GWs by using improved
analysis methods~\cite{Hashino:2018wee,Cline:2021iff}. However, we do not consider
these methods in our analysis for the sake of simplicity.

As shown in Fig.~\ref{fig:GWs}, the peak height of the GW amplitude
lowers as the value of $\beta_{Q}$ gets smaller in
general. Here, one has to mention that for small $\upsilon_{S}$ values,
the light dilaton scenario is most likely to be detectable than the
PRHM one. The qualitatively reason is as follows. The strength of
the first order phase transition is generally related to the potential
difference at the zero temperature between the symmetric vacuum
$(H,s)=(0,0)$ and the broken vacuum $(H,s)=(\upsilon,\upsilon_{S})$.
The potential difference is approximately given by
\begin{align}
V_{{\rm eff}}(0,0)-V_{{\rm eff}}(\upsilon,\upsilon_{S})\simeq\frac{N_{Q}m_{Q}^{4}}{128\pi^{2}},\label{eq:dVeff}
\end{align}
where the loop corrections from the SM particles are neglected here.
According to Eq.~\eqref{eq:dVeff}, the parameters
$N_{Q},\,\alpha_{Q}$ and $\beta_{Q}$ should be small to make
a high potential barrier, which means that the EWPT is
strongly first order in the parameter space region with small $N_{Q},\,\alpha_{Q}$
and $\beta_{Q}$.

Generally, in models with singlet scalar fields a non-trivial vacuum ($0,s\neq0$) can be preferred rather than the origin ($0,0$).
If such vacuum exists, the strongly first order EWPT can
be easily realized~\cite{Ahriche:2007jp}. In order to avoid the
existence of such non-trivial vacuum, the following condition
should be satisfied
\begin{equation}
\lambda_{S}+\delta\lambda_{S}+\frac{3}{32\pi^{2}}N_{Q}\beta_{Q}^{2}\left[\log\left(\frac{T^{2}}{\Lambda^{2}}\right)+0.99\right]>0.\label{eq:wrongvac}
\end{equation}
Since our model has a $Z_{2}$ symmetry $s\to-s$
even at the quantum level, the condition in Eq.~\eqref{eq:wrongvac}
is always satisfied.
Therefore, the approximation in Eq.~\eqref{eq:dVeff} is meaningful in our model.

We note that the typical values of $\alpha$ and $\widetilde{\beta}$ we found are around $0.1-1$ and $100-10000$, respectively.
Is is well known that the bubble collision contribution can be dominant when $\alpha \gg 1$~\cite{Caprini:2015zlo}.
Our results indicate that the dominant source of the GWs in our scenario is the sound wave contribution.

\begin{figure}[t]
\includegraphics[width=0.99\textwidth]{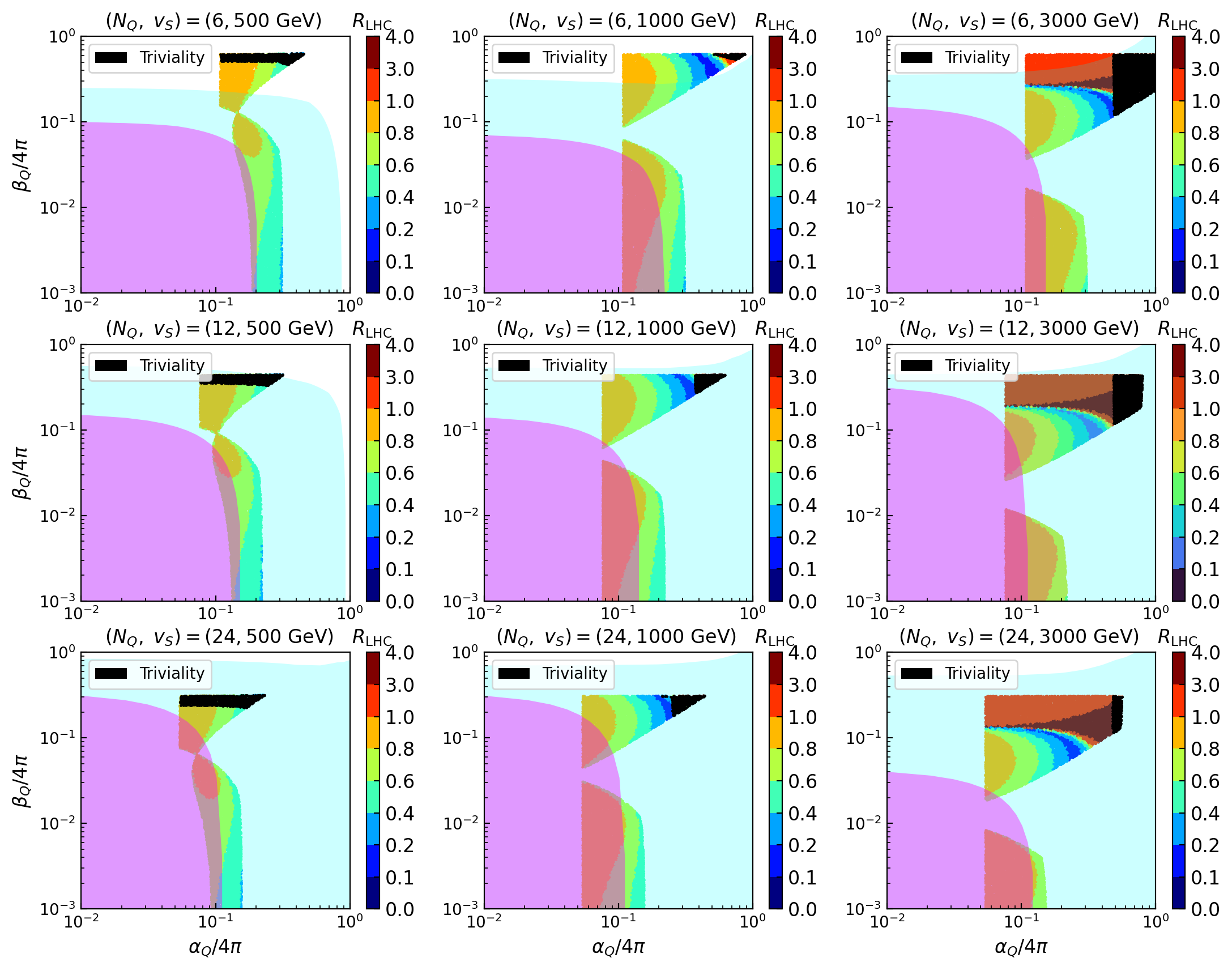} \caption{The parameter dependence for the value of $R_{{\rm LHC}}$. The cyan
region can be explored by the GW observations at the DECIGO. The
magenta region is excluded by the completion condition for the phase
transition in Eq~\eqref{eq:TransComp}. The large deviation in the
cross section for the di-Higgs production prefers the large $\alpha_{Q}$. The black region is constrained by the triviality bound with $\Lambda_{\rm Lan} < 10\,{\rm TeV}$.}
\label{fig:Rratio_LHC}
\end{figure}

\begin{figure}[ht]
\includegraphics[width=0.99\textwidth]{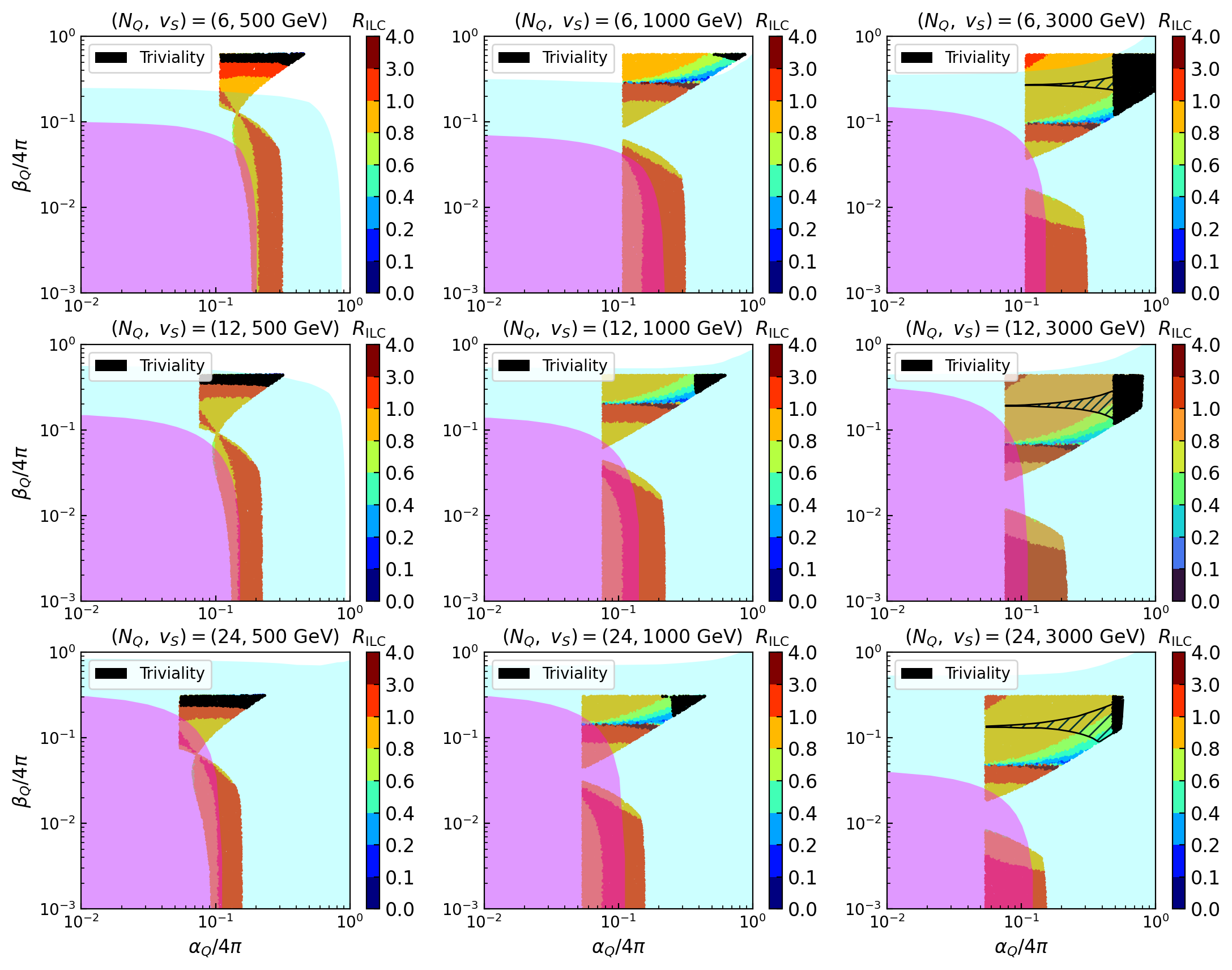} \caption{The parameter dependence for the value of $R_{{\rm ILC}}$. The cyan
region can be explored by the GW observations at the DECIGO. The
magenta region is excluded by the completion condition for the phase
transition in Eq.~\eqref{eq:TransComp}. The black shaded region
is the current constraint on the di-Higgs production from the LHC
results. The black region is constrained by the triviality bound with $\Lambda_{\rm Lan} < 10\,{\rm TeV}$.}
\label{fig:Rratio_ILC}
\end{figure}

As noted above, the parameter space region satisfying the condition in Eq.~\eqref{eq:vcTc}
may imply a large deviation in the cross section for
the di-Higgs production from the SM prediction. To confirm this, we
present the di-Higgs production cross section at the LHC and ILC in
Fig.~\ref{fig:Rratio_LHC} and Fig.~\ref{fig:Rratio_ILC}, respectively.
These figures show the ratio of $R_{{\rm LHC}}$ and $R_{{\rm ILC}}$
as defined in Eq.~\eqref{R} for different values of $N_{Q}$ and
$\upsilon_{S}$. These figures are produced by taking into account
the different theoretical and experimental constraints that are mentioned above.
According to Fig.~\ref{fig:Rratio_LHC} and Fig.~\ref{fig:Rratio_ILC},
the large $\alpha_{Q}$ values are preferred to realize a large deviation in the cross
section of the di-Higgs production from the SM prediction. As was
mentioned earlier, for the light dilaton case, the $R$ ratios cannot
be large due to the constraints from the mixing angle and the completion
condition for the phase transition. This means that we can distinguish
the PRHM and the light dilaton scenarios by combining the measurement
of the $R_{{\rm LHC}}$ and $R_{{\rm ILC}}$ with the GW observations.
By comparing Fig.~\ref{fig:Rratio_LHC} with Fig.~\ref{fig:Rratio_ILC},
the value of $R_{{\rm ILC}}$ is larger than $R_{{\rm LHC}}$. One
has to mention that the recent LHC negative searches for the di-Higgs
signal established the upper bound on the di-Higgs production cross
section $\sigma_{hh}^{{\rm LHC}}<112\,{\rm fb}$~\cite{CMS:2022dwd},
which excludes significant regions in the parameter space, especially
in the case with $\upsilon_{s}=3\,{\rm TeV}$. These regions are presented
in Fig.~\ref{fig:Rratio_ILC} as the black dashed regions. Thus,
it is expected that the ILC may impose more severe constraints on
the parameter space region with a large deviation in the $R$ ratios.

\begin{figure}[t]
\includegraphics[width=0.99\textwidth]{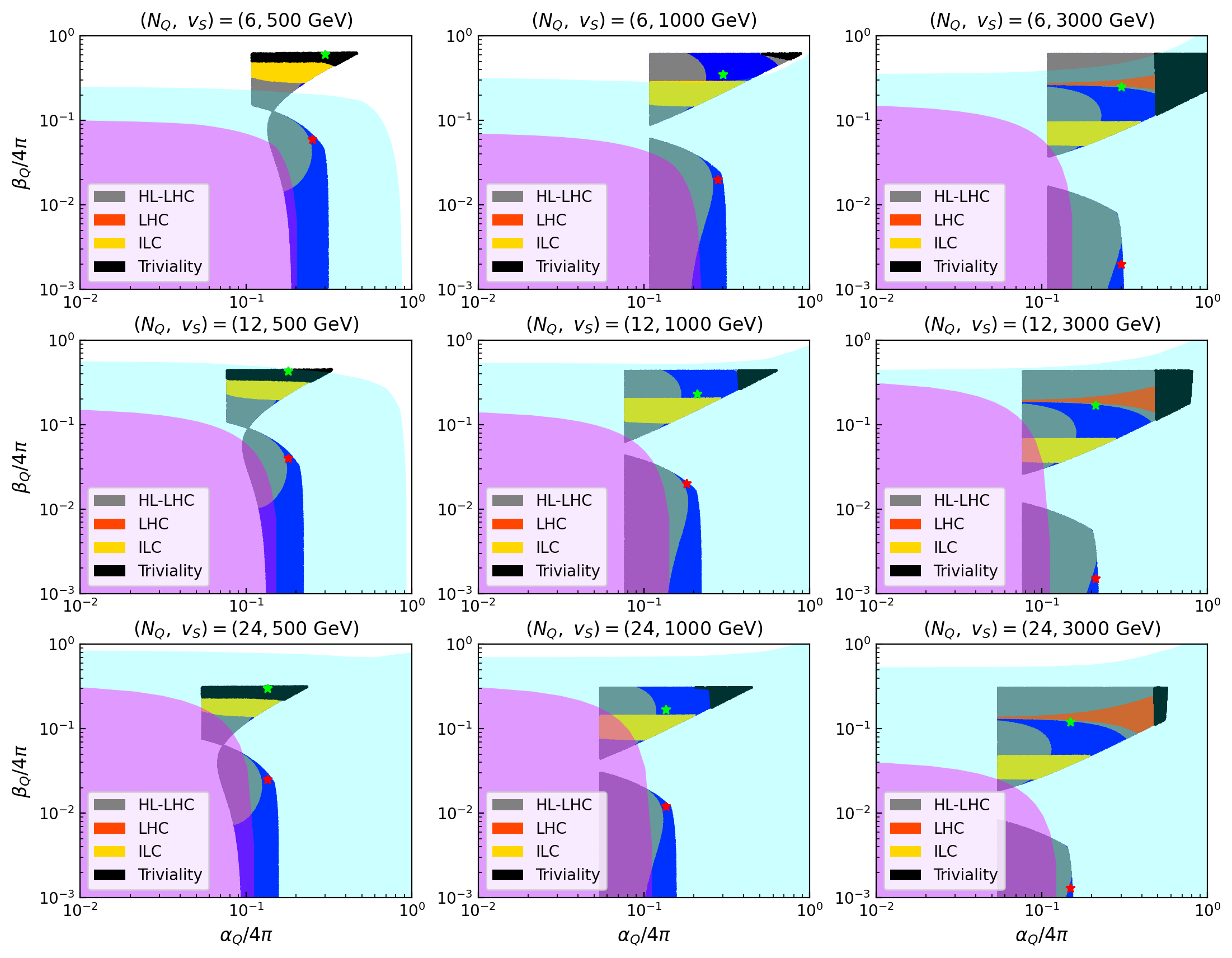}
\caption{The parameter space region explored by the LHC, HL-LHC and ILC. The gray,
orange and green regions are explored by HL-LHC, LHC and ILC, respectively.
DECIGO can explore the cyan shaded region. The magenta region is constrained
by the completion condition for the phase transition. The blue regions
are the unexplored parameter space regions.
The black region is constrained by the triviality bound with $\Lambda_{\rm Lan} < 10\,{\rm TeV}$.}
\label{fig:summary}
\end{figure}

In Fig.~\ref{fig:summary}, we show that the constraints on the model
from current and future collider experiments such as the LHC, HL-LHC
and ILC. The orange and gray regions can be explored by the LHC~\cite{CMS:2022dwd}
and HL-LHC~\cite{Cepeda:2019klc}, respectively. The yellow regions
are within the reach of the ILC with $\sqrt{s}=500\,{\rm GeV}$~\cite{Bambade:2019fyw},
while the blue regions cannot be explored by the di-Higgs production
measurement at the LHC, HL-LHC and ILC. On the other hand, the GW observations
can be used to probe the models within the blue region.

Even if the PRHM scenario cannot be distinguished from the dilaton
scenario by the di-Higgs production measurements, we can determine
which scenario is preferred by combining collider and GW results.
For instance, we focus on the two BPs with $N_{Q}=12$ and $v_{S}=3\,{\rm TeV}$ shown in Fig.~\ref{fig:Tc_Tn}.
The predicted GW spectra of these BPs has been shown in the panel with $N_{Q}=12$ and $v_{S}=3\,{\rm TeV}$ in Fig.~\ref{fig:GWs}.
One remarks that the positions of peak of the GWs are close.
However, the GW peak position in the PRHM scenario is different
from that in the light dilaton scenario. This implies that we are
able to determine which scenario is preferred by observing the GW
spectra even if a large cross section di-Higgs signal would not be
observed at future collider experiments. Therefore, one can determine
whether the PRHM scenario or the light dilaton scenario are realized
by utilizing complementarity of collider experiments and GW observations.

We provide a commentary on the Landau pole scale in our scenario. It has been numerically confirmed that the typical Landau pole scale in our model is around $O(1)$ TeV when using the beta functions outlined in Appendix~\ref{sec:Landau}. In Table~\ref{table:Landau_pole}, the Landau pole scales for the two BPs considered in Fig.~\ref{fig:GWs}, with mass-dependent beta functions, are presented. The results in Table~\ref{table:Landau_pole} indicate that the typical values of $\Lambda_{\text{\rm Lan}}$ in our model are around $O(10)$ TeV. This finding suggests that our phenomenological analyses satisfy perturbativity.

For the PRHM scenarios with $v_{s} = 500,{\rm GeV}$, the Landau pole scale can be in the $O(1)$ TeV range. However, such BPs are constrained by the triviality bound with $\Lambda_{\text{\rm Lan}}< 10,{\rm TeV}$.

\begin{table}[t]
\centering
\begin{tabular}{|c|c|c|c|c|c|}
\hline
$N_{Q}$ & $v_{S}$ = 500 GeV & $v_{S}= $1 TeV & $v_{S}=$3 TeV \\
\hline
6 & (R) 31.0 TeV (G) 7.05 TeV & (R) 19.8 TeV (G) 26.0 TeV & (R) 14.9 TeV (G) 53.0 TeV \\
\hline
12 & (R) 34.0 TeV (G) 6.66 TeV & (R) 37.6 TeV (G) 40.3 TeV & (R) 17.4 TeV (G) 92.8 TeV \\
\hline
24 & (R) 27.5 TeV (G) 6.19 TeV & (R) 29.7 TeV (G) 46.6 TeV & (R) 19.3 TeV (G) 96.9 TeV \\
\hline
\end{tabular}
\caption{The Landau pole scale values with the mass dependent beta functions. Here,
(R) and (G) denote the values for the BPs considered in Fig.~\ref{fig:GWs}.}
\label{table:Landau_pole}
\end{table}

Before closing this section, we would like to comment on the possibility of
projecting our results to other renormalizable new physics models.
In this work, we considered a generic CSI model, where the quantum effects
that trigger the EWSB lead to the EWPT dynamics and GWs discussed
above. These quantum corrections lead to the same behavior whether they are
a consequence of single/multiple field contributions that are just bosonic or bosonic
and fermionic contributions. However, in renormalizable new physics models,
additional model-dependent theoretical and experimental constraints should
be considered; and the viable parameter space is significantly affected.
For instance, if such a new physics model suggests a DM candidate; and/or
proposes a solution to the neutrino mass smallness problem, the new
masses and couplings in the model would be severely constrained not only
by the motivation requirements (DM relic density and/or neutrino mass smallness);
but also by other constraints like lepton flavor violating processes,
Higgs invisible and di-photon decays. For instance, in the current
generic model, if the singlet $Q$ is considered to be $U(1)_{Y}$ charged,
a significant part of the parameter space $\alpha_{Q},\,\beta_{Q},\,N_{Q}$
and $\upsilon_{S}$ would be excluded by the bound from $h\to\gamma\gamma$.
Therefore, one expects that the parameter space regions where the light dilation
or PRHM scenarios can be realized are in general narrowed down when
specific new physics models are considered.

\section{Conclusion~\label{sec:conclusion}}

In this work, we have investigated the possibility of detectable
GWs that are produced during a strongly first order EWPT, within a class of models
with the CSI. In order to get model independent results, we have considered the SM
extended by a real scalar singlet to assist the EWSB, and an additional scalar field with multiplicity $N_{Q}$, and couplings $\alpha_{Q}$ and $\beta_{Q}$ that accounts for the
RCs. By scanning over the model free parameters ($N_{Q}$, $\alpha_{Q}$,
$\beta_{Q}$ and $\upsilon_{S}$), the impact of the RCs on the EWPT
and GW observables have been estimated. Especially, we have focused
on the difference between the light dilaton and PRHM scenarios. We
have analyzed the properties of the GWs from a strongly first order EWPT
by considering a generic model with the CSI.
We have shown that peaked GW spectrum might be detected at
future space-based interferometers such as the DECIGO and other
GW observations even if the condition $\upsilon_{c}/T_{c}>1$ is not satisfied.
As a result, the wide parameter space region in the CSI
model may be tested by utilizing the GW observations.

Also, we have analyzed the cross section in the di-Higgs production
processes at the LHC and ILC. As we have shown, the cross section
can be more than three times larger than the SM prediction in the
PRHM scenario. As shown in Fig.~\ref{fig:summary}, the parameter
regions with such large deviation in the di-Higgs production process
can be explored at the LHC, HL-LHC and ILC. In addition, we have shown
that GW observations are useful to distinguish the light dilaton
scenario from the PRHM scenario even if collider experiments cannot
observe the di-Higgs production signal. This fact indicates that
we may be able to determine whether the light dilaton scenario or
the PRHM scenario is preferred by utilizing complementary of collider
experiments and GW observations.

\section*{Acknowledgements}

A. A. was funded by the University of Sharjah under the research projects
No 21021430107 ``\textit{Hunting for New Physics at Colliders}''
and No 23021430135 ``\textit{Terascale Physics: Colliders vs Cosmology}''.
S. K. was supported by the JSPS KAKENHI Grant No.~20H00160, No.~202F21324 and No.~203K17691.
M. T. was supported by JSPS KAKENHI Grant No. JP21J10645. M. T. would like to thank Sharjah university (IAESTE exchange program, HEP group, RISE and VCRGS) for the financial support and the warm hospitality.

\appendix

\section{Renormalization Group Equations}~\label{sec:Landau}

In this appendix, the renormalization group equations for the couplings in our model are discussed.
The beta functions for each coupling in our model at one-loop level are given by
\begin{align}
&16 \pi^2 \beta_{g'} = \frac{41}{6} g'^3, 16 \pi^2 \beta_{g} = -\frac{19}{6} g^3, 16 \pi^2 \beta_{g_{3}} = -7 g_{3}^3, \nonumber \\
&16 \pi^2 \beta_{y_{t}} = y_{t} \left( - 8 g_{3}^2 - \frac{9}{4} g^2 - \frac{17}{12} g'^2 + \frac{9}{2} y_{t}^2 \right), \nonumber \\
&16 \pi^2 \beta_{\lambda_{h}} = 4 \lambda_{h}^2 + 3 N_{Q} \alpha_{Q}^2 + 3 \omega^2 + \frac{27}{4} g^4 + \frac{9}{2} g^2 g'^2 + \frac{9}{4} g'^4 - 36 y_{t}^4 - \lambda_{h} (9 g^2 + 3g'^2 - 12 y_{t}^2) , \nonumber \\
&16 \pi^2 \beta_{\lambda_{s}} = 3 \lambda_{s}^2 + 3 N_{Q} \beta_{Q}^2 + 12 \omega^2, 16 \pi^2 \beta_{\lambda_{Q}} = \frac{N_{Q} + 8}{3} \lambda_{Q}^2 + 3 \beta_{Q}^2 + 12 \alpha_{Q}^2 , \nonumber \\
&16 \pi^2 \beta_{\omega} = 4 \omega^2 + \omega \lambda_{s} + 2 \lambda_{h} \omega + N_{Q} \alpha_{Q} \beta_{Q} + 6 \omega y_{t}^2 - \frac{9}{2} \omega g^2 - \frac{3}{2} \omega g'^2, \nonumber \\
&16 \pi^2 \beta_{\alpha_{Q}} = 4 \alpha_{Q}^2 + \frac{N_{Q}+2}{3} \alpha_{Q} \lambda_{Q} + 2 \alpha_{Q} \lambda_{h} + \omega \beta_{Q} + 6 \alpha_{Q} y_{t}^2 - \frac{9}{2} \alpha_{Q} g^2 - \frac{3}{2} \alpha_{Q} g'^2, \nonumber \\
&16 \pi^2 \beta_{\beta_{Q}} = 4 \beta_{Q}^2 + \frac{N_{Q} + 2}{3} \lambda_{Q} \beta_{Q} + \lambda_{s} \beta_{Q} + 4 \omega \alpha_{Q},
\end{align}
where $g'$, $g$ and $g_{3}$ are gauge couplings for $U(1)_{Y}$, $SU(2)_{L}$ and $SU(3)_{C}$ gauge groups, respectively.
Also, $y_{t}$ is the Yukawa coupling for top quarks.
It is confirmed that these equations are consistent with the previous work if we take $N_{Q}=1$~\cite{Abada:2013pca}.


\begin{thebibliography}{1}

\bibitem{ATLAS:2012yve}
G.~Aad et al. [ATLAS],
Phys. Lett. B \textbf{716}, 1-29 (2012) [arXiv:1207.7214 [hep-ex]].
S.~Chatrchyan \textit{et al.} [CMS],
Phys. Lett. B \textbf{716} (2012), 30-61 [arXiv:1207.7235 [hep-ex]].


\bibitem{Bardeen:1995kv}
W.~A.~Bardeen,
FERMILAB-CONF-95-391-T.


\bibitem{Lee:2012jn}
J.~S.~Lee and A.~Pilaftsis,
Phys. Rev. D \textbf{86}, 035004 (2012) [arXiv:1201.4891 [hep-ph]].


\bibitem{Englert:2013gz}
C.~Englert, J.~Jaeckel, V.~V.~Khoze
and M.~Spannowsky,
JHEP \textbf{04}, 060 (2013) [arXiv:1301.4224 [hep-ph]].


\bibitem{Guo:2014bha}
J.~Guo and Z.~Kang,
Nucl. Phys. B \textbf{898}, 415-430 (2015) [arXiv:1401.5609 [hep-ph]].


\bibitem{Endo:2015ifa}
K.~Endo and Y.~Sumino,
JHEP \textbf{05}, 030 (2015) [arXiv:1503.02819 [hep-ph]].


\bibitem{Hashino:2015nxa}
K.~Hashino, S.~Kanemura and Y.~Orikasa,
Phys. Lett. B \textbf{752}, 217-220 (2016) [arXiv:1508.03245 [hep-ph]].

\bibitem{Hashino:2016rvx}
K.~Hashino, M.~Kakizaki, S.~Kanemura
and T.~Matsui,
Phys. Rev. D \textbf{94}, 015005 (2016) [arXiv:1604.02069 [hep-ph]].


\bibitem{Ahriche:2016cio}
A.~Ahriche, K.~L.~McDonald and S.~Nasri,
JHEP \textbf{06}, 182 (2016) [arXiv:1604.05569 [hep-ph]].


\bibitem{Ahriche:2016ixu}
A.~Ahriche, A.~Manning, K.~L.~McDonald
and S.~Nasri,
Phys. Rev. D \textbf{94}, no.5, 053005 (2016) [arXiv:1604.05995 [hep-ph]].


\bibitem{Lane:2019dbc}
K.~Lane and E.~Pilon,
Phys. Rev. D \textbf{101}, no.5, 055032 (2020) [arXiv:1909.02111
[hep-ph]].


\bibitem{Kanemura:2020yyr}
S.~Kanemura and M.~Tanaka,
Phys. Lett. B \textbf{809}, 135711 (2020) [arXiv:2005.05250 [hep-ph]].


\bibitem{Braathen:2020vwo}
J.~Braathen, S.~Kanemura and M.~Shimoda,
JHEP \textbf{03}, 297 (2021) [arXiv:2011.07580 [hep-ph]].

\bibitem{Ahriche:2021frb}
A.~Ahriche,
 Nucl. Phys. B \textbf{982} (2022), 115896 [arXiv:2110.10301 [hep-ph]].

\bibitem{Soualah:2021xbn}
R.~Soualah and A.~Ahriche,
Phys. Rev. D \textbf{105} (2022) no.5, 055017 [arXiv:2111.01121
[hep-ph]].

\bibitem{Eichten:2022vys}
E.~J.~Eichten and K.~Lane,
Phys. Rev. D \textbf{107}, no.7, 075038 (2023)
[arXiv:2209.06632 [hep-ph]].

\bibitem{Coleman:1973jx}
S.~R.~Coleman and E.~J.~Weinberg,
Phys. Rev. D \textbf{7}, 1888-1910 (1973).

\bibitem{Gildener:1976ih}
E.~Gildener and S.~Weinberg,
Phys. Rev. D \textbf{13} (1976), 3333.

\bibitem{Planck:2018vyg} N.~Aghanim \textit{et al.} [Planck],
Astron. Astrophys. \textbf{641} (2020), A6 [erratum: Astron. Astrophys.
\textbf{652} (2021), C4] [arXiv:1807.06209 [astro-ph.CO]].

\bibitem{Sakharov:1967dj}
A.~D.~Sakharov,
Pisma Zh. Eksp. Teor. Fiz. \textbf{5} (1967), 32-35

\bibitem{Kuzmin:1985mm}
V.~A.~Kuzmin, V.~A.~Rubakov and M.~E.~Shaposhnikov,
Phys. Lett. B \textbf{155} (1985), 36

\bibitem{Bochkarev:1987wf}
A.~I.~Bochkarev and M.~E.~Shaposhnikov,
Mod. Phys. Lett. A \textbf{2} (1987), 417.
A.~I.~Bochkarev, S.~V.~Kuzmin
and M.~E.~Shaposhnikov, 
 Phys. Rev. D \textbf{43} (1991), 369-374

\bibitem{Ahriche:2007jp}
A.~Ahriche,
Phys. Rev. D \textbf{75} (2007), 083522 [arXiv:hep-ph/0701192 [hep-ph]].


\bibitem{Ahriche:2014jna}
A.~Ahriche, T.~A.~Chowdhury and S.~Nasri,
JHEP \textbf{11} (2014), 096 [arXiv:1409.4086 [hep-ph]].

\bibitem{Fuyuto:2014yia}
K.~Fuyuto and E.~Senaha,
Phys. Rev. D \textbf{90}, no.1, 015015 (2014) [arXiv:1406.0433 [hep-ph]].

\bibitem{Kanemura:2022ozv}
S.~Kanemura and M.~Tanaka,
Phys. Rev. D \textbf{106}, no.3, 035012 (2022) [arXiv:2201.04791
[hep-ph]].

\bibitem{Kanemura:2004ch}
S.~Kanemura, Y.~Okada and E.~Senaha,
Phys. Lett. B \textbf{606}, 361-366 (2005) [arXiv:hep-ph/0411354
[hep-ph]].

\bibitem{Shifman:1979eb}
M.~A.~Shifman, A.~I.~Vainshtein, M.~B.~Voloshin
and V.~I.~Zakharov,
Sov. J. Nucl. Phys. \textbf{30}, 711-716 (1979) ITEP-42-1979.

\bibitem{Kanemura:2002vm}
S.~Kanemura, S.~Kiyoura, Y.~Okada, E.~Senaha
and C.~P.~Yuan,
Phys. Lett. B \textbf{558}, 157-164 (2003) [arXiv:hep-ph/0211308
[hep-ph]].

\bibitem{Kanemura:2004mg}
S.~Kanemura, Y.~Okada, E.~Senaha and
C.~P.~Yuan,
Phys. Rev. D \textbf{70}, 115002 (2004) [arXiv:hep-ph/0408364 [hep-ph]].

\bibitem{Arhrib:2012ia}
A.~Arhrib, R.~Benbrik and N.~Gaur,
Phys. Rev. D \textbf{85}, 095021 (2012)
[arXiv:1201.2644 [hep-ph]].

\bibitem{Kanemura:2019kjg}
S.~Kanemura, M.~Kikuchi, K.~Mawatari, K.~Sakurai and K.~Yagyu,
Nucl. Phys. B \textbf{949}, 114791 (2019)
[arXiv:1906.10070 [hep-ph]].

\bibitem{Kanemura:2019slf}
S.~Kanemura, M.~Kikuchi, K.~Mawatari, K.~Sakurai and K.~Yagyu,
Comput. Phys. Commun. \textbf{257}, 107512 (2020)
[arXiv:1910.12769 [hep-ph]].

\bibitem{Degrassi:2023eii}
G.~Degrassi and P.~Slavich,
[arXiv:2307.02476 [hep-ph]].

\bibitem{Aiko:2023nqj}
M.~Aiko, J.~Braathen and S.~Kanemura,
[arXiv:2307.14976 [hep-ph]].

\bibitem{Kosowsky:1991ua}
A.~Kosowsky, M.~S.~Turner and R.~Watkins,
Phys. Rev. D \textbf{45} (1992), 4514-4535.

\bibitem{Kamionkowski:1993fg}
M.~Kamionkowski, A.~Kosowsky and
M.~S.~Turner,
Phys. Rev. D \textbf{49} (1994), 2837-2851 [arXiv:astro-ph/9310044
[astro-ph]].

\bibitem{Schwaller:2015tja}
P.~Schwaller,
Phys. Rev. Lett. \textbf{115} (2015) no.18, 181101 [arXiv:1504.07263
[hep-ph]].

\bibitem{Kosowsky:1992rz}
A.~Kosowsky, M.~S.~Turner and R.~Watkins,
Phys. Rev. Lett. \textbf{69} (1992), 2026-2029.

\bibitem{Grojean:2006bp}
C.~Grojean and G.~Servant,
 Phys.\ Rev.\ D \textbf{75}, 043507 (2007) [hep-ph/0607107].

\bibitem{LISA:2017pwj}
P.~Amaro-Seoane \textit{et al.} [LISA],
[arXiv:1702.00786 [astro-ph.IM]].

\bibitem{Kawamura:2006up}
S.~Kawamura, T.~Nakamura, M.~Ando, N.~Seto,
K.~Tsubono, K.~Numata, R.~Takahashi, S.~Nagano, T.~Ishikawa and
M.~Musha, \textit{et al.}
 Class. Quant. Grav. \textbf{23} (2006), S125-S132.

\bibitem{Kakizaki:2015wua}
M.~Kakizaki, S.~Kanemura and T.~Matsui,
Phys. Rev. D \textbf{92}, no.11, 115007 (2015)
[arXiv:1509.08394 [hep-ph]].

\bibitem{Huang:2016cjm}
P.~Huang, A.~J.~Long and L.~T.~Wang,
Phys. Rev. D \textbf{94}, no.7, 075008 (2016)
[arXiv:1608.06619 [hep-ph]].

\bibitem{Hashino:2016xoj}
K.~Hashino, M.~Kakizaki, S.~Kanemura, P.~Ko and T.~Matsui,
Phys. Lett. B \textbf{766}, 49-54 (2017)
[arXiv:1609.00297 [hep-ph]].

\bibitem{Hashino:2018zsi}
K.~Hashino, M.~Kakizaki, S.~Kanemura, P.~Ko and T.~Matsui,
JHEP \textbf{06}, 088 (2018)
[arXiv:1802.02947 [hep-ph]].


\bibitem{Brdar:2018num}
V.~Brdar, A.~J.~Helmboldt and J.~Kubo,
JCAP \textbf{02}, 021 (2019)
[arXiv:1810.12306 [hep-ph]].

\bibitem{Brdar:2019qut}
V.~Brdar, A.~J.~Helmboldt and M.~Lindner,
JHEP \textbf{12}, 158 (2019)
[arXiv:1910.13460 [hep-ph]].

\bibitem{Salvio:2023qgb}
A.~Salvio,
JCAP \textbf{04}, 051 (2023)
[arXiv:2302.10212 [hep-ph]].

\bibitem{Salvio:2023ynn}
A.~Salvio,
[arXiv:2307.04694 [hep-ph]].


\bibitem{Farzinnia:2014yqa}
A.~Farzinnia and J.~Ren,
Phys. Rev. D \textbf{90} (2014) no.7, 075012 [arXiv:1408.3533 [hep-ph]].

\bibitem{Mohamadnejad:2019vzg}
A.~Mohamadnejad,
Eur. Phys. J. C \textbf{80} (2020) no.3, 197 [arXiv:1907.08899 [hep-ph]].

\bibitem{Kubo:2015joa}
J.~Kubo and M.~Yamada,
PTEP \textbf{2015} (2015) no.9, 093B01 [arXiv:1506.06460 [hep-ph]].

\bibitem{Sannino:2015wka}
F.~Sannino and J.~Virkajervi,
Phys. Rev. D \textbf{92} (2015) no.4, 045015 [arXiv:1505.05872 [hep-ph]].

\bibitem{Kierkla:2022odc}
M.~Kierkla, A.~Karam and B.~Swiezewska,
JHEP \textbf{03}, 007 (2023)
[arXiv:2210.07075 [astro-ph.CO]].

\bibitem{Hashino:2021qoq}
K.~Hashino, S.~Kanemura and T.~Takahashi,
Phys. Lett. B \textbf{833}, 137261 (2022) [arXiv:2111.13099 [hep-ph]].

\bibitem{Hashino:2022tcs}
K.~Hashino, S.~Kanemura, T.~Takahashi
and M.~Tanaka, 
Phys. Lett. B \textbf{838}, 137688 (2023) [arXiv:2211.16225 [hep-ph]].

\bibitem{Subaru}
https://hsc.mtk.nao.ac.jp/ssp/.

\bibitem{OGLE}
http://ogle.astrouw.edu.pl.

\bibitem{PRIME}
http://www-ir.ess.sci.osaka-u.ac.jp/prime/index.html

\bibitem{Roman}
https://roman.gsfc.nasa.gov


\bibitem{Alexander-Nunneley:2010tyr}
L.~Alexander-Nunneley and A.~Pilaftsis,
JHEP \textbf{09}, 021 (2010)
[arXiv:1006.5916 [hep-ph]].

\bibitem{Kanemura:2012hr}
S.~Kanemura, E.~Senaha, T.~Shindou and T.~Yamada,
JHEP \textbf{05}, 066 (2013)
[arXiv:1211.5883 [hep-ph]].



\bibitem{CMS:2022dwd}
A.~Tumasyan \textit{et al.} [CMS], 
Nature \textbf{607}, no.7917, 60-68 (2022) [arXiv:2207.00043 [hep-ex]].

\bibitem{Cepeda:2019klc}
M.~Cepeda, S.~Gori, P.~Ilten, M.~Kado,
F.~Riva, R.~Abdul Khalek, A.~Aboubrahim, J.~Alimena, S.~Alioli
and A.~Alves, \textit{et al.} 
CERN Yellow Rep. Monogr. \textbf{7}, 221-584 (2019) [arXiv:1902.00134
[hep-ph]].

\bibitem{Bambade:2019fyw}
P.~Bambade, T.~Barklow, T.~Behnke, M.~Berggren,
J.~Brau, P.~Burrows, D.~Denisov, A.~Faus-Golfe, B.~Foster and
K.~Fujii, \textit{et al.} 
[arXiv:1903.01629 [hep-ex]].

\bibitem{Arhrib:2009hc}
A.~Arhrib, R.~Benbrik, C.~H.~Chen, R.~Guedes and R.~Santos,
JHEP \textbf{08}, 035 (2009)
[arXiv:0906.0387 [hep-ph]].

\bibitem{Dolan:2012ac}
M.~J.~Dolan, C.~Englert and M.~Spannowsky,
Phys. Rev. D \textbf{87}, no.5, 055002 (2013)
[arXiv:1210.8166 [hep-ph]].

\bibitem{Kanemura:2016lkz}
S.~Kanemura, M.~Kikuchi and K.~Yagyu,
Nucl. Phys. B \textbf{917}, 154-177 (2017)
[arXiv:1608.01582 [hep-ph]].

\bibitem{Dawson:2017jja}
S.~Dawson and M.~Sullivan,
Phys. Rev. D \textbf{97}, no.1, 015022 (2018)
[arXiv:1711.06683 [hep-ph]].

\bibitem{Carena:2018vpt}
M.~Carena, Z.~Liu and M.~Riembau,
Phys. Rev. D \textbf{97}, no.9, 095032 (2018)
[arXiv:1801.00794 [hep-ph]].

\bibitem{Arco:2021bvf}
F.~Arco, S.~Heinemeyer and M.~J.~Herrero,
Eur. Phys. J. C \textbf{81}, no.10, 913 (2021)
[arXiv:2106.11105 [hep-ph]].

\bibitem{Abouabid:2021yvw}
H.~Abouabid, A.~Arhrib, D.~Azevedo, J.~E.~Falaki, P.~M.~Ferreira, M.~M\"uhlleitner and R.~Santos,
JHEP \textbf{09}, 011 (2022)
[arXiv:2112.12515 [hep-ph]].

\bibitem{Iguro:2022fel}
S.~Iguro, T.~Kitahara, Y.~Omura and H.~Zhang,
Phys. Rev. D \textbf{107}, no.7, 075017 (2023)
[arXiv:2211.00011 [hep-ph]].


\bibitem{ATLAS:2016neq}
G.~Aad \textit{et al.} [ATLAS and CMS],
JHEP \textbf{08} (2016), 045 [arXiv:1606.02266 [hep-ex]].


\bibitem{Lindner:1985uk}
M.~Lindner,
Z. Phys. C \textbf{31}, 295 (1986)

\bibitem{Kanemura:2023wap}
S.~Kanemura and Y.~Mura,
[arXiv:2310.15622 [hep-ph]].


\bibitem{Grojean:2004xa}
C.~Grojean, G.~Servant and J.~D.~Wells,
Phys. Rev. D \textbf{71}, 036001 (2005) [arXiv:hep-ph/0407019 [hep-ph]].


\bibitem{Noble:2007kk}
A.~Noble and M.~Perelstein, 
Phys. Rev. D \textbf{78} (2008), 063518 [arXiv:0711.3018 [hep-ph]].

\bibitem{Ahriche:2014cpa}
A.~Ahriche, A.~Arhrib and S.~Nasri,
Phys. Lett. B \textbf{743} (2015), 279-283 [arXiv:1407.5283 [hep-ph]].

\bibitem{Spira:1995mt}
M.~Spira,
[arXiv:hep-ph/9510347 [hep-ph]].

\bibitem{ATLAS:2020zms}
G.~Aad \textit{et al.} [ATLAS],
Phys. Rev. Lett. \textbf{125} (2020) no.5, 051801 [arXiv:2002.12223
[hep-ex]].

\bibitem{ATLAS:2020tlo}
G.~Aad \textit{et al.} [ATLAS],
 Eur. Phys. J. C \textbf{81} (2021) no.4, 332 [arXiv:2009.14791
[hep-ex]].

\bibitem{CMS:2021klu}
A.~Tumasyan \textit{et al.} [CMS],
[arXiv:2109.06055 [hep-ex]].

\bibitem{ATLAS:2021fet}
G.~Aad \textit{et al.} [ATLAS],
ATLAS-CONF-2021-030.

\bibitem{ATLAS:2021ulo}
G.~Aad \textit{et al.} [ATLAS],
ATLAS-CONF-2021-035.

\bibitem{ATLAS:2021jki}
G.~Aad \textit{et al.} [ATLAS],
 ATLAS-CONF-2021-016.


\bibitem{Croon:2020cgk}
D.~Croon, O.~Gould, P.~Schicho, T.~V.~I.~Tenkanen and G.~White,
JHEP \textbf{04}, 055 (2021)
[arXiv:2009.10080 [hep-ph]].

\bibitem{Lofgren:2021ogg}
J.~L\"ofgren, M.~J.~Ramsey-Musolf, P.~Schicho and T.~V.~I.~Tenkanen,
Phys. Rev. Lett. \textbf{130}, no.25, 251801 (2023)
[arXiv:2112.05472 [hep-ph]].

\bibitem{Schicho:2022wty}
P.~Schicho, T.~Tenkanen, V.I. and G.~White,
JHEP \textbf{11}, 047 (2022)
[arXiv:2203.04284 [hep-ph]].


\bibitem{Dolan:1973qd}
L.~Dolan and R.~Jackiw,
Phys.\ Rev.\ D \textbf{9}, 3320 (1974).
Weinberg, 
Phys.\ Rev.\ D \textbf{9},3357 (1974).

\bibitem{Carrington:1991hz}
M.~E.~Carrington,
Phys. Rev. D \textbf{45}, 2933 (1992).

\bibitem{Gross:1980br}
D.~J.~Gross, R.~D.~Pisarski and L.~G.~Yaffe,
Rev. Mod. Phys. \textbf{53} (1981), 43.

\bibitem{Ahriche:2018rao}
A.~Ahriche, K.~Hashino, S.~Kanemura
and S.~Nasri,
 Phys. Lett. B \textbf{789} (2019), 119-126 [arXiv:1809.09883 [hep-ph]].

\bibitem{Linde:1981zj}
A.~D.~Linde,
Nucl. Phys. B \textbf{216}, 421 (1983) [erratum: Nucl. Phys. B \textbf{223},
544 (1983)].

\bibitem{Wainwright:2011kj}
C.~L.~Wainwright,
Comput. Phys. Commun. \textbf{183}, 2006-2013 (2012) [arXiv:1109.4189
[hep-ph]].

\bibitem{Huber:2008hg}
S.~J.~Huber and T.~Konstandin,
JCAP \textbf{09} (2008), 022 [arXiv:0806.1828 [hep-ph]].

\bibitem{Caprini:2006jb}
C.~Caprini and R.~Durrer,
Phys. Rev. D \textbf{74} (2006), 063521 [arXiv:astro-ph/0603476
[astro-ph]].

\bibitem{Caprini:2015zlo}
C.~Caprini, M.~Hindmarsh, S.~Huber,
T.~Konstandin, J.~Kozaczuk, G.~Nardini, J.~M.~No, A.~Petiteau,
P.~Schwaller and G.~Servant,~\textit{et al.}
JCAP \textbf{04} (2016), 001. [arXiv:1512.06239 [astro-ph.CO]].

\bibitem{Espinosa:2010hh}
J.~R.~Espinosa, T.~Konstandin, J.~M.~No
and G.~Servant,
JCAP \textbf{06}, 028 (2010) [arXiv:1004.4187 [hep-ph]].

\bibitem{Khojali:2022squ}
M.~O.~Khojali, A.~Abdalgabar, A.~Ahriche and A.~S.~Cornell,
Phys. Rev. D \textbf{106} (2022) no.9, 095039 
[arXiv:2206.06211 [hep-ph]].
\bibitem{Baouche:2021wwa}
N.~Baouche, A.~Ahriche, G.~Faisel and S.~Nasri,
Phys. Rev. D \textbf{104} (2021) no.7, 075022 
[arXiv:2105.14387 [hep-ph]].

\bibitem{TianQin:2015yph}
J.~Luo \textit{et al.} [TianQin],
Class. Quant. Grav. \textbf{33}, no.3, 035010 (2016) [arXiv:1512.02076
[astro-ph.IM]].

\bibitem{Ruan:2018tsw}
W.~H.~Ruan, Z.~K.~Guo, R.~G.~Cai and
Y.~Z.~Zhang,
Int. J. Mod. Phys. A \textbf{35}, no.17, 2050075 (2020) [arXiv:1807.09495
[gr-qc]].

\bibitem{Hashino:2018wee}
K.~Hashino, R.~Jinno, M.~Kakizaki, S.~Kanemura,
T.~Takahashi and M.~Takimoto,
Phys. Rev. D \textbf{99}, no.7, 075011 (2019) [arXiv:1809.04994
[hep-ph]].


\bibitem{Cline:2021iff}
J.~M.~Cline, A.~Friedlander, D.~M.~He,
K.~Kainulainen, B.~Laurent and D.~Tucker-Smith,
Phys. Rev. D \textbf{103}, no.12, 123529 (2021) [arXiv:2102.12490
[hep-ph]].


\bibitem{Abada:2013pca}
A.~Abada and S.~Nasri,
Phys. Rev. D \textbf{88}, no.1, 016006 (2013)
[arXiv:1304.3917 [hep-ph]].



\end{thebibliography}
\end{document}